\documentclass[pra,12pt,showpacs,onecolumn,superscriptaddress,longbibliography]{revtex4-1}

\usepackage{graphicx}
\usepackage{color}
\usepackage{dcolumn}
\usepackage[caption=false]{subfig}
\newcommand\Tstrut{\rule{0pt}{2.6ex}}         
\newcommand\Bstrut{\rule[-0.9ex]{0pt}{0pt}}   
\usepackage[flushleft]{threeparttable}
\usepackage{array}
\usepackage{adjustbox}
\usepackage{hyperref} 
\usepackage{soul}

\graphicspath{{./Figures_new/}}

\begin{document}

\title{Emulating optical cycling centers in polyatomic molecules}

\author{Ming Li}
\affiliation{Department of Physics, Temple University, Philadelphia, Pennsylvania 19122, USA}
\author{Jacek K{\l}os}
\affiliation{Department of Physics, Temple University, Philadelphia, Pennsylvania 19122, USA}
\affiliation{Department of Chemistry and Biochemistry, University of Maryland, College Park, Maryland 20742, USA}
\author{Alexander Petrov}
\affiliation{Department of Physics, Temple University, Philadelphia, Pennsylvania 19122, USA}
\affiliation{NRC, Kurchatov Institute PNPI, Gatchina, 188300, and Division of Quantum Mechanics, St.~Petersburg State University, St.~Petersburg, 199034, Russia}
\author{Svetlana Kotochigova}
\email{skotoch@temple.edu}
\affiliation{Department of Physics, Temple University, Philadelphia, Pennsylvania 19122, USA}

\date{\today}

\begin{abstract}
An optical cycling center (OCC) is a recently coined term to 
indicate two electronic states within a complex quantum object 
that can repeatedly experience optical laser excitation and 
spontaneous decay, while being well isolated from its environment. 
Here we present a quantitative understanding of electronic, 
vibrational, and rotational excitations of the polyatomic
SrOH molecule, which possesses a localized OCC near its Sr atom. 
In particular, we describe the vibrationally-dependent trends 
in the Franck-Condon factors of the bending and stretching modes 
of the molecular electronic states coupled in the optical transition.
These simulations required us to perform electronic structure
calculations of the multi-dimensional potential energy surfaces 
of both ground and excited states, the determination 
of vibrational and bending modes, and corresponding 
Franck-Condon factors. We also discuss the extent to which 
the optical cycling center has diagonal Franck-Condon factors.
\end{abstract}

\maketitle

\newpage

Laser cooling and trapping of atoms, enabled by the existence 
of closed optical cycling transitions, have revolutionized 
atomic physics and led to breakthroughs in several disciplines 
of science and technology \cite{Wieman1999}. 
These advances enabled the simulation of exotic phases in 
quantum-degenerate atomic gases, the creation of a novel
generation of atomic clocks, matter-wave interferometry, 
and the development of other highly-sensitive sensors.  
Temperatures below tens of microkelvin have also allowed 
the confinement of diatomic molecules, built from or associated 
with laser-cooled atoms, in electric, magnetic, and/or 
optical traps, where they are isolated from their environment 
and can be carefully studied.

Achieving similar temperatures for polyatomic molecules, however,
remains challenging. Since polyatomic molecules are characterized 
by multiple degrees of freedom and have correspondingly more complex
structures, it is far from obvious whether there exists polyatomic 
molecules that have the nearly-closed optical cycling transitions 
required for successful laser cooling. Such transitions could then 
repeatedly scatter photons.

A diverse list of promising applications for ultracold 
polyatomic molecules exists. This includes creating novel types 
of sensors, advancing quantum information science, simulation of 
complex exotic materials, performing precision spectroscopy to 
test the Standard Model of particle physics, and, excitingly, 
the promise of control of quantum chemical reactions when 
each molecule is prepared in a unique rovibrational quantum state.  
Moreover, the de Broglie wavelength of colliding ultracold molecules 
is much larger than the range of intermolecular forces
and, thus, the science of the breaking and making of chemical bonds 
has entered into an unexplored regime.

Over the decades many spectroscopic studies of molecules 
consisting of an alkaline-earth metal atom (M) and a ligand have been 
performed \cite{Presunka:1995,WORMSBECHER1,WORMSBECHER2,Hilborn1983,
BERNATH1984, Brazier1987,Bopegedera1987,Brazier1989,BERNATH1991},
predominantly to determine their structure. 
The simplest polyatomic molecules of this type, the triatomic 
alkaline-earth monohydroxides M-OH, have attracted particular 
attention after the discovery of their peculiar ionic chemical bond. 
When a ground-state alkaline-earth atom is bound to OH one of 
its two outer-most $ns^2$ electrons is transferred to the ligand, 
leaving the second electron in an open shell molecular orbital 
localized around the metal atom. This electron can be optically 
excited without disturbing the atom-ligand bond leading to 
so-called highly-diagonal FCFs and efficient optical cycling. High-resolution laser-spectroscopy experiments on CaOH (and CaOD) \cite{Hilborn1983} and SrOH (and SrOD) \cite{Nakagawa1983} were first to deduce the strong ionic character of the metal-hydroxide bond. Reference \cite{Langhoff1998} showed that the remaining valence electron of the metal atom can be easily promoted to any excited orbital.  

Recently, Doppler laser cooling \cite{Doppler1985} of the simpler 
CaF, SrF,  YbF, and YO dimers has been demonstrated
\cite{DeMille2016,Doyle2,Hinds2017,Collopy2018,Lim2018}. 
Their success is associated with nearly-diagonal Franck-Condon 
factors (FCFs) on an allowed optical molecular transition. 
Diagonal FCFs ensure that spontaneous emission 
from vibrational state $v$ of the excited electronic state
populates with near unit probability vibrational state $v'=v$ 
 of the ground electronic state.
This closed two-level system can absorb
and emit thousands of photons per molecule to achieve cooling.

Following these successful molecular cooling experiments,
Isaev and Berger \cite{Isaev2016} suggested that
similar near-diagonality of FCFs exists 
in other classes of polyatomic molecules.  
It is now understood that metal-monohydroxides and even larger 
metal-monoalkoxide molecules with metal atom as Sr, Ca, or Ba are 
promising candidates for laser cooling.
In 2017 the first optical cycling transitions in the 
polyatomic monohydroxide molecule SrOH  was demonstrated by 
Dr.~Doyle's group \cite{Doyle2017}. 
They succeeded in reducing the translational motion of SrOH to 
the record low temperature of 750 microkelvin
starting from 50 milikelvin and using the near unit values of the FCFs.
To eliminate rotational branching during the photon cycling process experimentalists 
use of the rotationally closed $J^{\prime\prime} \to J^{\prime} -1$ angular momentum transitions (see for details Refs.~\cite{Doyle2017,Kosyryev_2019}).

A significant effort from the scientific community, however, 
is needed to identify and study the classes of polyatomic molecules 
with closed optical cycling transitions. Laser cooling of polyatomic molecules is relevant to those molecules that are able to scatter hundreds of photons (stimulated absorption followed by a spontaneous emission) between two vibrational states without loss to other vibrational states, i.e. have a cycling transition where for each cycle the kinetic energy of the molecule is reduced by  $\Delta E/k=1$ $\mu$K and 10 $\mu$K depending on the cooling process and $k$ is the Boltzmann constant.  The requirement of 100 scattering cycles is to a certain degree arbitrary but reasonable and  implies that FCF larger than 0.99 are needed. A larger FCF will allow better cooling.

In fact, a comprehensive analysis of level structures that 
are amenable to laser cooling has to be conducted. 
Special attention must be given to finding excited-state molecular potentials that have: 
i) a shape that is similar to that of the ground-state potential; ii) a strong dipole electronic transition with
the ground state; and iii) weak transitions to dark states that are not optically coupled to the excited state. 
It is to be expected that only a small number of polyatomic molecules have these three properties.

An engineering approach, however, can be used to add optical cycling centers (OCCs) to a variety of 
polyatomic molecules with increasing complexity \cite{Klos2019}. For this approach to work we need to ensure 
that the electrons holding the optically active atom and the molecule together remain sufficiently localized so that the whole molecular system can be translationally cooled. This stringent requirement motivates our intent to conduct an assessment of the role of electron density and localization in the coupling to the ligand molecule. Recent  theoretical studies on the cooling properties of alkaline-earth monohydroxides other than SrOH
can be found in Refs.~\cite{Isaev2016,Kozyryev2016}. They only determined molecular parameters and Frank-Condon 
factors for vibronic transitions between ground and excited states  near  equilibrium geometries.

Our objectives are to better quantify the extent to which polyatomic SrOH is an ideal platform for cooling with  laser light
as well as to determine the  ``global'' shape of its ground and excited potential energy surfaces (PES) in anticipation of future 
research on dissociating SrOH into Sr and OH and on ultracold scattering between Sr and OH. In this paper, we describe a complex electronic structure determination of local and some global properties of four multi-dimensional intramolecular PESs 
of SrOH as well as calculate their corresponding vibrational structures.  Our effort involves locating potential minima, 
potential avoided crossings as well as conical intersections (CIs), where two adiabatic PESs of the same electronic 
symmetry touch. We represent the PESs and thus the CIs both in terms of  adiabatic and diabatic representations. We also 
determine  Frank-Condon factors for not-only the lowest vibrational states but also  higher-excited 
vibrational states of the  potentials, where the transitions are no longer diagonal due to non-adiabatic and anharmonic
corrections. At near linear geometries, we evaluate the Renner-Teller (RT) parameter \cite{RennerTeller} for
some excited-state potentials.

\vspace*{6mm}

\noindent
{\large \bf Results}\\
\noindent
{\bf Electronic potentials of SrOH}. The computation of PESs is crucial for defining the landscape 
in which the nuclei transverse upon interacting with one another. 
These surfaces can have many features in the context of their
topology that are important for the internal transition mechanisms.
We begin by describing our calculation of the ground and excited 
PESs of strontium monohydroxide SrOH. Past theoretical studies on 
M-OH were devoted to either spectroscopic characteristics near 
equilibrium geometries \cite{Vaeck2017} or the PESs for the 
lighter BeOH and MgOH with fixed O-H separation \cite{Pesalakis1999}. Experimental studies of CaOH \cite{BERNATH1991} and SrOH
\cite{Presunka:1995,Brazier1987,Bopegedera1987}  
predominantly focussed on their near-equilibrium 
ro-vibrational structure. 

\begin{figure}
\includegraphics[width=0.35\textwidth,trim=0 10 0 0,clip]{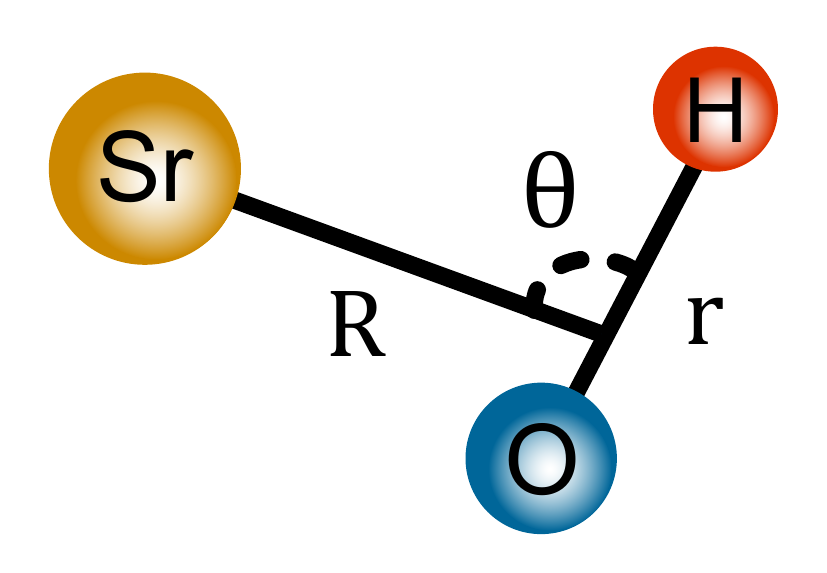}
\caption{
Illustration of the relevant Jacobi coordinates {\bf R} and 
{\bf r} for SrOH. Here, $R$ is the separation between Sr and
the center of mass of OH, $r$ is the separation between O and H, 
and $\theta$ is the angle between vectors {\bf R} and {\bf r}.
 }
\label{fig:Jacobi}
\end{figure}

\begin{figure*}
\includegraphics[scale=0.48,trim=15 80 0 0,clip]{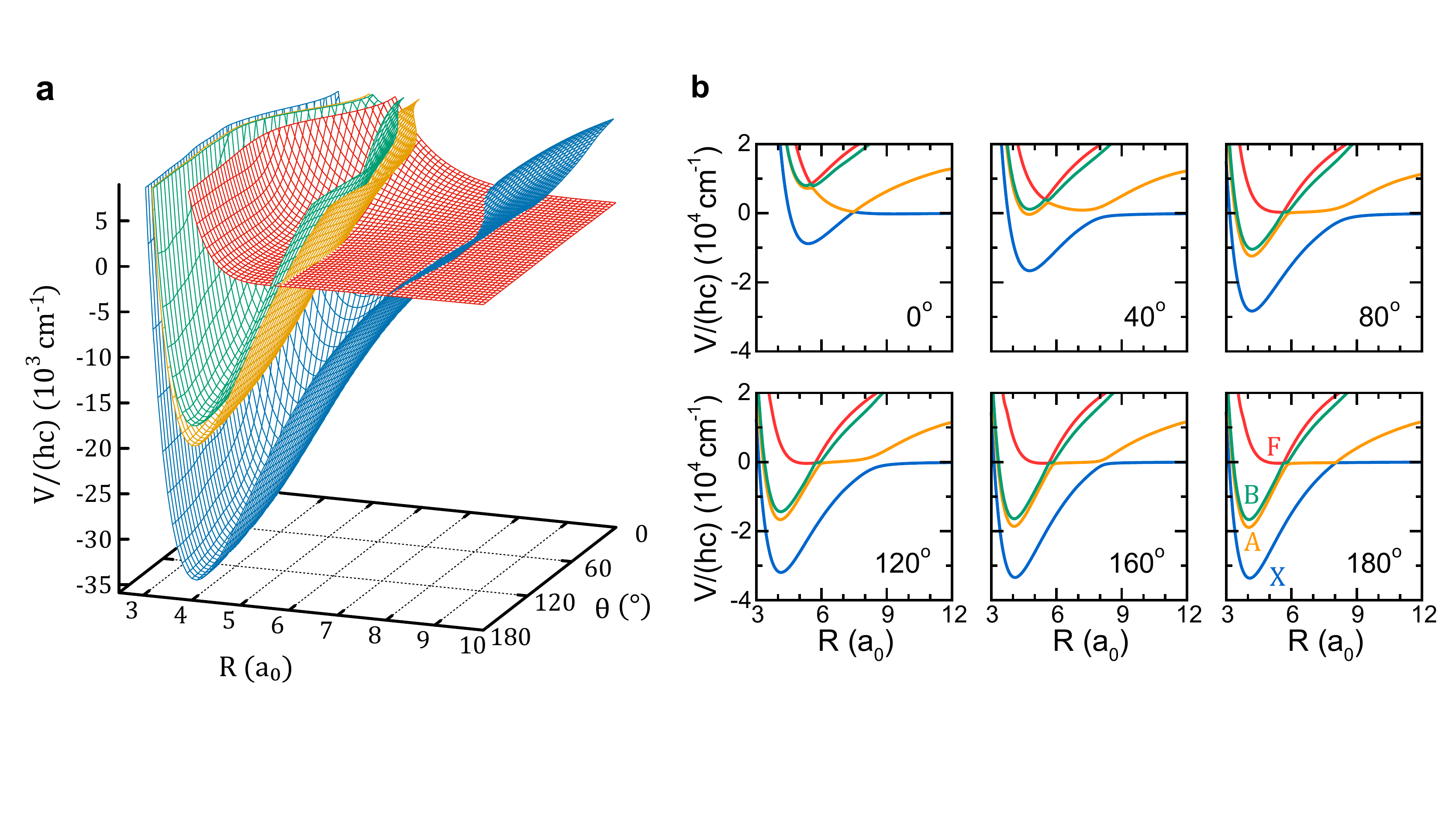}
\caption{The ground and excited potential energy surfaces $V(R,\theta)$ of SrOH 
at separation ${r=1.832a_0}$.
a) A two-dimensional cut through the four relevant $^2\!A^{\prime}$ 
diabatic potential surfaces as functions of separation $R$ and angle $\theta$. The blue, orange, green, and red curves 
correspond to states 
$1\, ^2\!A^{\prime} ({\rm X} ^2\Sigma^+)$, 
$2\, ^2\!A^{\prime} ({\rm A}^2\Pi)$, 
$3\, ^2\!A^{\prime} ({\rm B}^2\Sigma^+)$, and 
$4\, ^2\!A^{\prime}({\rm F}^2\Pi)$, respectively. 
Seams, containing conical intersections at the two linear geometries, 
are lines along which two diabatic potentials are degenerate. 
b) Adiabatic $^2\!A^{\prime}$ potentials as functions of separation $R$ 
for six values of angle $\theta$. The absolute ground state 
minimum occurs at angle ${\theta=180^{\circ}}$ at separation $R\approx 4a_0$.
The zero of energy is set at the energy of a ground-state Sr atom infinitely far away from the 
ground-state OH(X$^2\Pi$) molecule at its equilibrium bond length.  
Potential energies are expressed in units of cm$^{-1}$, 
using Planck's constant $h$ and speed of light in vacuum $c$. 
}
\label{fig:SrOH_pots}
\end{figure*}

The relevant multi-dimensional PESs have been determined using 
a combination of coupled-cluster (CCSD(T)), equation-of-motion 
coupled-cluster (EOM-CC), and multi-reference configuration 
interaction (MRCI) methods. This combination allows us to overcome 
the limitations of the CCSD(T) and EOM-CC methods associated with 
their single reference nature and the MRCI method with its limits 
on the size of active space and characterize multiple avoided crossings 
between relevant PESs. Since the molecule contains one heavy atom,
relativistic effects are embedded into the description of the 
core electrons. Potentials are presented in the Jacobi coordinates 
{\bf R} and {\bf r} defined in Fig.~\ref{fig:Jacobi}. 
For our purposes it is sufficient to determine the PESs, 
which are only functions of $(R,\theta,r)$ in the two-dimensional 
plane with the OH separation fixed to its diatomic equilibrium 
separation as the energy required to 
vibrationally excite OH is nearly seven times larger that than those of the Sr-OH stretch and bend. 
Coupling to the OH stretch can then be ignored for our objectives.
(We always use ${r=1.832a_0}$ 
and also suppress the $r$ dependence in our notation.)

The PESs and corresponding electronic wavefunctions are labeled by 
their total electronic spin angular momentum, 
here always a doublet or spin 1/2, as well as the irreducible 
representations of point groups $C_{\infty v}$ for linear geometries
and $C_s$ otherwise. In particular, we obtain four 
$^2\!A^{\prime}$ and two $^2\!A^{\prime\prime}$ potentials 
using standard notation for the irreducible representations of 
$C_s$, respectively.
For linear geometries $^2\!A^{\prime}$ states connect to 
either $^2\Sigma^+$ or $^2\Pi$ states of the $C_{\infty v}$ 
symmetry group, while $^2\!A^{\prime\prime}$ states always 
become $^2\Pi$ states. Hence, we label potential surfaces by 
$n\,^2\!A^{\prime,\prime\prime}(m \,{^2\!\Lambda^\pm})$,
where $n=1,2,3,\dots$ and $m={\rm X, A, B,} \dots$ indicate 
the energy ordering of states for $R$ and $r$ close to 
their equilibrium separations and $\theta=180^{\circ}$.

CCSD(T) and EOM-CC calculations lead to the most-accurate 
PESs for each irreducible representation and, in principle, 
should return the corresponding adiabatic PES with 
avoided crossings from excited potentials in the same 
irreducible representations. Numerically, however, we find this 
not to be true due to the fact that the method use only  
excitations from a single reference configuration and that 
the electronic wavefunctions rapidly change from ionic to 
covalently bonded with small changes in the Jacobi coordinates, 
in particular regions of $R$. In essence, for a given reference 
configuration the CCSD(T) or EOM-CC simulations return diabatic
potentials with electronic wavefunctions that have either an ionic 
bond, where one of the outer valence electrons of Sr is now tightly 
bound to oxygen, or a covalent bond, where both valence electrons 
mostly remain in orbit around the Sr nucleus and barely couple with 
the OH electron cloud. The diabatic PESs have lines in the 
plane $(R,\theta)$ along which two potentials have the same energy.

Figure~\ref{fig:SrOH_pots}a shows four diabatic 
PESs of SrOH as functions of $R$ and $\theta$ obtained by CCSD(T) 
and EOM-CC methods.  Details regarding electronic basis sets, coupled-cluster 
method for the individual $^2\!A^{\prime}$ and $^2\!A^{\prime\prime}$ PESs,  and interpolation can be found in Methods.
The calculations are performed at ten angular and about 45 radial geometries. 
This data is then interpolated using an analytical functional form. 
The interpolated PESs are essential in recognizing system 
characteristics, such as minima and transition states, i.\,e. 
saddle points, as well as crossings. 
Near extrema the curvature or Hessian matrix is evaluated. 

The optimized geometry for the ground-state SrOH molecule occurs 
at $\theta = 180^{\circ}$. Its electronic wavefunction has 
X$^2\Sigma^+$ symmetry and is ionically bonded. These observations 
are consistent with spectroscopic data \cite{Presunka:1994} 
and the semi-empirical analyses of Ref.~\cite{Nguyen:2018}. 
The $2\,^2\!A^{\prime}$ and $3\,^2\!A^{\prime}$ states also have 
ionic character and minima at $\theta = 180^{\circ}$. 
In fact, the minima of these three potentials have 
nearly the same  equilibrium coordinates and curvatures providing 
excellent conditions for optical cycling. These conditions are 
further described in subsection ``Wavefunction overlaps''.

The fourth diabatic PES is the shallow excited 
$4\, ^2\!A^{\prime}({\rm F}^2\Pi)$ potential with a covalently 
bonded electronic state that dissociates to Sr($^1$S) and 
OH(X$^2\Pi$). For large $R$ the ionic PESs correlate to 
electronically-excited states of the Sr atom. 
Crucially, we observe that the three ionic diabatic PESs 
each have a curved line in the $(R,\theta)$ plane along which  
the $4\, ^2\!A^{\prime}({\rm F}^2\Pi)$ potential equals the energy 
of the ionic potential. It is worth noting that to good approximation
this curve is independent of $\theta$. 
The $4\, ^2\!A^{\prime}({\rm F}^2\Pi)$ potential is expected to 
play a crucial role in the zero-energy dissociation of SrOH to create
ultracold OH fragments, an important radical for various scientific
applications.

Figure~\ref{fig:SrOH_pots}b shows the interpolated 
adiabatic $^2\!A'$ PESs of the SrOH molecule as functions of 
$R$ for six values of $\theta$ and ${r=1.832a_0}$ obtained by 
diagonalizing the $4\times4$ Hamiltonian with the diabatic 
PESs as diagonal matrix elements and coupling matrix elements 
described in subsection ``Non-adiabatic couplings and conical intersections'' at each $(R,\theta)$. 
These cuts through the PESs exhibit several intriguing features. 
First, for $\theta=0^\circ$ and $180^\circ$ CIs, 
where two potentials still touch, are apparent.  
For other values of $\theta$ the crossings are avoided.
The location of the CIs at linear geometries is specific to SrOH, other molecules
will have CIs at other geometries.
CIs lead to geometric or Berry phases \cite{Berry:1984}, 
i.\,e. sign changes associated with an electronic wavefunction 
when transported along a closed path encircling a CI. 
These phases lead to interference that allows efficient 
non-adiabatic transitions between surfaces \cite{Guo:2017}, 
modify product rotational state distributions in chemical reactions,
and influence ro-vibrationally averaged transition matrix elements.

We have similarly obtained two diabatic
$^2\!A^{\prime\prime}$ PESs. Both are $^2\Pi$ states at linear 
$C_{\infty v}$ geometries. On the energy scale of 
Fig.~\ref{fig:SrOH_pots}a their shapes are nearly indistinguishable 
from those of the  $2\, ^2\!A^{\prime}({\rm A}^2\Pi)$ and 
$4\, ^2\!A^{\prime}({\rm F}^2\Pi)$ potentials. In fact, 
$1\, ^2\!A^{\prime\prime}({\rm A}^2\Pi)$ and 
$2\, ^2\!A^{\prime}({\rm A}^2\Pi)$ as well as 
$2\, ^2\!A^{\prime\prime}({\rm F}^2\Pi)$ and 
$4\, ^2\!A^{\prime}({\rm F}^2\Pi)$ are degenerate for linear geometries.
Unlike the corresponding $^2\!A^\prime$ potentials, however, 
the two $^2\!A^{\prime\prime}$ potentials do not possess CIs.

\vspace*{6mm}

\noindent
{\bf Non-adiabatic couplings and conical intersections}. The adiabatic $^2\!A^{\prime}$ PESs 
in Fig.~\ref{fig:SrOH_pots}b are Born-Oppenheimer (BO) 
potentials \cite{BO} that are typically used as a starting point in describing scattering and chemical processes.
The BO approximation is based on the realization that the motion of nuclei and electrons occur
on different time or energy scales and usually only a single PES is required. This leads to a significant simplification 
of the description of scattering and chemical processes.

For SrOH the BO approximation breaks down when potentials of 
$^2\!A^\prime$ symmetry come close and even become degenerate 
for one or more geometries.  
Couplings between these potentials can not then be neglected. 
The corresponding non-adiabatic transition probability for 
``hopping'' between potentials increases dramatically especially 
for conical intersections and can greatly affect chemical
properties \cite{Domcke2004}. 

\begin{figure*}
\includegraphics[width=0.9\textwidth,trim=0 0 0 0,clip]{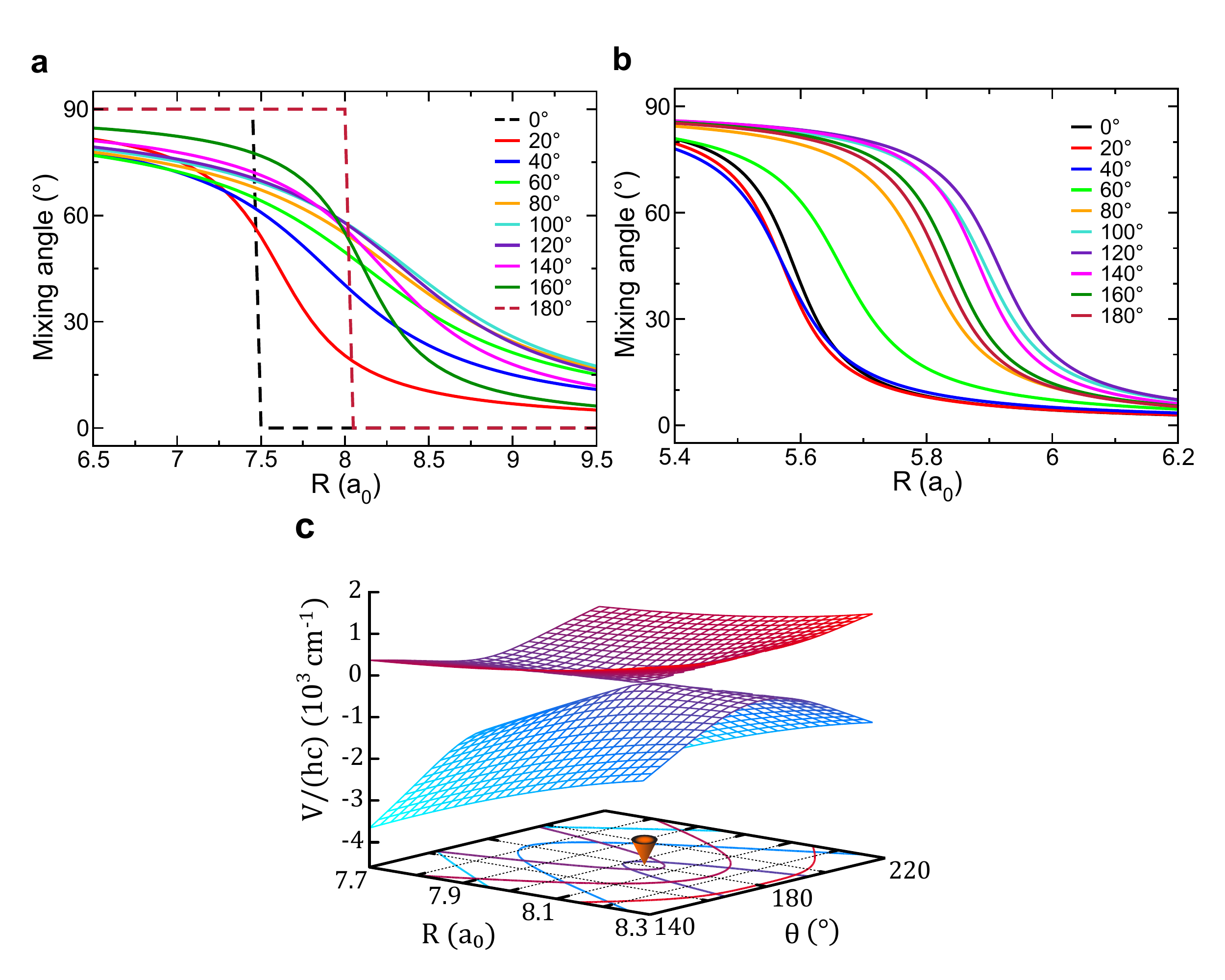}
\caption{Characterization of mixing angles and conical intersections between the $1\,^2\!A^{\prime}$  and $4\,^2\!A^{\prime}$ states.
a) Mixing angles $\vartheta(R,\theta)$ for the diabatic 
$1\,^2\!A^{\prime}$  and $4\,^2\!A^{\prime}$ states as functions 
of separation $R$ for ten different values of angle $\theta$ determined 
from our quantum-mechanical calculations. Conical intersections are apparent as the $-90^\circ$ 
jump in mixing angle for  curves with $\theta=0^\circ$ and $180^\circ$.
b) Mixing angles similarly determined from our calculations for the 
$1\,^2\!A^{\prime\prime}$ and $2\,^2\!A^{\prime\prime}$ diabatic states. 
No conical intersection is present.
c) Adiabatic $1\,^2\!A^{\prime}$ (blue surface) and 
$4\,^2\!A^{\prime}$ (red surface) potential energy surfaces $V$ as functions of separation $R$ and 
and angle $\theta$ near the conical intersection  at  $R\approx 8a_0$ and  $\theta=180^\circ$. 
} 
\label{fig:SrOH_mixing}
\end{figure*}

Our CCSD(T) and EOM-CC calculations only resulted in diabatic PESs 
$V_{m}(R,\theta)$ and electronic wavefunctions $|\phi_{m}\rangle$, 
due to their reliance on a single reference configuration. 
Here, index $m$ labels states and corresponding PESs.
As the diabatic electronic wavefunctions are predominantly 
described by this single reference configuration, their dependence 
on $R$ and $\theta$ is weak and negligible, and suppressed in 
our notation. 

Coupling matrix elements between the diabatic potentials are constructed by 
performing less-accurate MRCI calculations at the 
SD level with a basis set similar to that used for our 
CCSD(T) calculations (details are given in Methods.)  
Our MRCI calculations rely on two or more reference configurations
and thus do lead to adiabatic BO potentials $U_{n}(R,\theta)$ 
and adiabatic electronic wavefunctions 
$|\psi_n(R,\theta)\rangle$, where index $n$ labels states. 
Adiabatic wavefunctions strongly depend on $R$ and $\theta$ 
near crossings or avoid crossings between potentials. 
We also compute diagonal overlap matrix elements 
$\langle\psi_n(R',\theta')|\psi_n(R,\theta)\rangle$ 
at different geometries within the MRCI method.

Numerically, we find that the general shapes of the CCSD(T) and MRCI 
potentials are the same. Specifically, the geometries where 
diabatic potentials in Fig.~\ref{fig:SrOH_pots}a cross and 
adiabatic MRCI potentials avoid are in good agreement. 
We can then assume that away from the crossings
$|\psi_n(R,\theta)\rangle\approx|\phi_m\rangle$ for some diabatic 
index $m$. Near avoided crossings the adiabatic wavefunction is 
a superposition of diabatic wavefunctions.
 
We focus on the coupling matrix element between the diabatic 
$1\,^2\!A^{\prime}$ and $4\,^2\!A^{\prime}$ states near $R=8a_0$ 
as well as those between $1\,^2\!A^{\prime\prime}$ and 
$2\,^2\!A^{\prime\prime}$ near $R=6a_0$. 
We assume the coupling between the diabatic 
$2\,^2\!A^{\prime}$ and $4\,^2\!A^{\prime}$ states near $R=6a_0$
is the same as between the $A^{\prime\prime}$ ones.
For each pair of states 
the adiabatic wavefunctions can then be approximated as the
superposition \cite{Nikitin2006Drake}
\begin{equation}
 \left( \begin{array}{c}|\psi_n(R,\theta)\rangle \\ 
 |\psi_{n'}(R,\theta)\rangle \end{array}\right) =
  \left(\begin{array}{cc} \cos\vartheta(R,\theta) 
  &\sin\vartheta(R,\theta)\\
 -\sin\vartheta(R,\theta) 
  & \cos\vartheta(R,\theta)
  \end{array}
  \right)
 \left(\begin{array}{c}|\phi_{m}\rangle \\ 
 |\phi_{m'}\rangle\end{array}
 \right) 
\end{equation}
with indices $n,n'$ and $m,m'$, respectively, where mixing angle 
$\vartheta(R,\theta)$ is to be determined.

To determine the mixing angle we, first, identify a reference geometry
$R_{\rm ref}$ for each $\theta$ away from the crossings, where 
$|\psi_n(R_{\rm ref},\theta)\rangle\approx|\phi_m\rangle$, 
$|\psi_{n'}(R_{\rm ref},\theta)\rangle\approx|\phi_{m'}\rangle$, 
and we can assume $\vartheta(R_{\rm ref},\theta)=0$. 
The mixing angle at other radial geometries is then given by
\begin{equation}
\vartheta(R,\theta)=\arccos \left[\langle\psi_n(R_{\rm ref},  
                    \theta)|\psi_n(R,\theta)\rangle \right]
 \label{eq:mixang}
\end{equation}
and the coupling matrix element between the two diabatic states 
$m$ and $m'$ is
\begin{equation}
W_{mm'}(R,\theta) =
  \frac{1}{2}[V_{m}(R,\theta)-V_{m'}(R,\theta)] 
  \tan [2\vartheta (R,\theta)]  \;.
\label{eq:H12}
\end{equation}
In the limit $V_{m}(R,\theta)\to V_{m'}(R,\theta)$ this expression 
needs to be treated carefully. The mixing angle will approach $\pi/4$ 
and only at CIs $W_{mm'}(R,\theta) =0$.

In practice, we parametrize $\vartheta(R,\theta)$ as we perform MRCI calculations only on a restricted set of geometries $(R_i,\theta_j)$.
We ensure a smooth functional dependence on $R$ at fixed $\theta$ 
by using
\begin{equation}
\vartheta(R,\theta)=-\frac{1}{2}\mathrm{arctan}  
   \left[\frac{R-R_c(\theta)}{R_0(\theta)}\right]
   +\frac{\pi}{4} \;,
\label{eq:mixangmod}
\end{equation}
where $R_{c}(\theta)$ is the curve where 
$V_{m}(R,\theta)= V_{m'}(R,\theta)$ and $R_0(\theta)\ge0$ is a 
coupling width that is fitted to reproduce the MRCI overlap matrix at  
$\theta_j$ (at other $\theta$ it is found with the Akima 
interpolation method~\cite{AkimaIn}.) By construction $R_0(\theta)$ 
is much smaller than $|R_{\rm ref}-R_{c}(\theta)|$.
We use $R_{\rm ref} = 11a_0$ and $7a_0$ for the pair of 
$^2\!A^{\prime}$ and $^2\!A^{\prime\prime}$ states, respectively.
Values for $\theta_j$, $R_{c}$, and $R_0$ are available upon request.

Figures~\ref{fig:SrOH_mixing}a and b show mixing angles 
$\vartheta(R,\theta)$ for the two diabatic $^2\!A^{\prime}$ 
and the two $^2\!A^{\prime\prime}$ states, respectively.
In both panels the mixing angle is seen to change relatively rapidly 
over a small range of $R$, i.\,e. determined by $R_0(\theta)$, 
from nearly $90^\circ$ to $0^\circ$. 
When $\vartheta(R,\theta)=45^\circ$ we have $R=R_c(\theta)$.
For the $^2\!A^{\prime}$ states in panel a, $R_0(\theta)$
is largest for $\theta \approx 90^\circ$ and is zero for 
$\theta\to0^\circ$ and $180^\circ$, indicative of CIs
where $\vartheta(R,\theta)$ has an infinitely sharp jump from 
$90^\circ$ to $0^\circ$.  For the $^2\!A^{\prime\prime}$ states 
in panel b the width $R_0(\theta)$ is nearly independent of $\theta$
and there is no CI. Finally, Fig.~\ref{fig:SrOH_mixing}c shows the 
adiabatic $^2\!A^{\prime}$ potentials near the CI at 
$R\approx8a_0$ and $\theta=180^\circ$ as determined by diagonalizing 
the $2\times2$ matrix containing the relevant diabatic PESs 
coupled by the coupling matrix element at each $(R,\theta)$. 
The figure corresponds to a surface plot of the data shown in 
Fig.~\ref{fig:SrOH_pots}b.

\begin{figure}
  \includegraphics[width=0.5\textwidth,trim=60 30 50 40,clip]{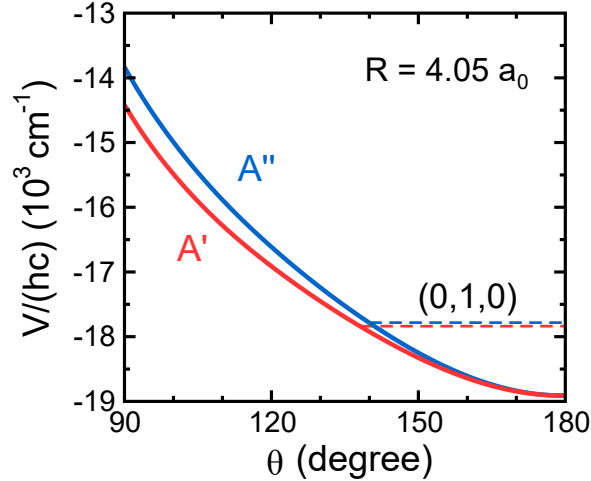}
 \caption{
 The splitting of $2\,^2\!A^{\prime}$ and $1\,^2\!A^{\prime \prime}$ potential energy surfaces $V$ as functions of angle $\theta$ at the 
 separations $R=4.05a_0$ and ${r=1.832a_0}$. The potentials are degenerate at $\theta = 180^{\circ}$.
 Dashed lines indicate the calculated energies of the first excited bending modes 
 $(0,1,0)$ for the two potentials, respectively.}
 \label{fig:A2Pi_splitting}
\end{figure}

The  Renner-Teller effect ~\cite{Renner1934,Teller1933} occurs in linear or quasi-linear, open-shell triatomic molecules and relies on 
non-adiabatic coupling between rovibrational and electronic motion in polyatomic molecules. 
It is associated with an energy degeneracy of electronic states for high-symmetry geometries 
that is lifted by vibrational bending motion. The motion causes degenerate potential energy 
surfaces to split and breaks the Born-Oppenheimer (BO) approximation. Stretching modes, 
which do not break linear symmetry, are not affected. 

We calculate the RT parameter by estimating the non-adiabatic coupling between 
$2\,^2\!A^\prime$ and  $1\,^2\!A^{\prime\prime}$ PESs. These PESs are degenerate at 
co-linear geometries but split otherwise.

Figure~\ref{fig:A2Pi_splitting} shows the potentials near the 
co-linear equilibrium separation $R$ as functions of the bending 
angle $\theta$. The potentials are degenerate at $\theta=180^{\circ}$. 
The slight anisotropy away from linear geometry causes the bending vibrational motion 
in the two PESs to differ. Our calculated $(0,1,0)$ bending energy relative to $(0,0,0)$ for the $2\,^2\!A^\prime$ and 
$1\,^2\!A^{\prime\prime}$ states are $\omega(A^{\prime})=384$ cm$^{-1}$ 
and $\omega(A^{\prime\prime})=412$ cm$^{-1}$, respectively, 
corresponding to a 28 cm$^{-1}$ difference. 
In the harmonic approximation this differential splitting is 
characterized by the Renner-Teller parameter \cite{Presunka:1994} 
\begin{equation}
\epsilon=
 \frac{\omega(A^{\prime})^2-\omega(A^{\prime\prime})^2}
      {\omega(A^{\prime})^2+\omega(A^{\prime\prime})^2} = -0.0693\;.
\end{equation}
Presunka \& Coxon~\cite{Presunka:1994} estimated a Renner-Teller
parameter of $-0.0791$ for these states from fitting this and other 
non-adiabatic parameters to spectroscopic data. 
Here, we calculate $\epsilon$ for the first time from {\em ab-initio}
calculations and find agreement with spectroscopic estimate to within 15\%.

\vspace* {6mm}

\noindent
{\bf Wavefunction overlaps}. Laser cooling of SrOH \cite{Doyle2017} relied on optical cycling 
transitions between either  
${\rm X}^2\Sigma^+ \leftrightarrow {\rm A}^2\Pi$ or
${\rm X}^2\Sigma^+ \leftrightarrow {\rm B}^2\Sigma^+$ vibronic states. 
The success of the experiment is associated with near-diagonal 
FCFs between the sets of levels supported by 
these electronic potentials, which ensures that spontaneous emission 
from a rovibrational state of the excited electronic state populates 
with near unit probability the corresponding rovibrational state of 
the ground electronic state. This two-level system absorbs and emits 
many photons to achieve cooling \cite{Doppler1985}.

Table~\ref{MO_composition} shows atomic orbital composition for the ground and A and B electronically excited states of SrOH. The X electronic ground state is mostly described by Sr-$s$ orbital as expected.  One can notice  that in the case of  the A and B states there is some participation of $d$-orbitals of Sr atom. For the A state that participation is quite small and the excited state is mainly formed by Sr-$p$ orbital participating in $s-p$ transition. The B state has quite a substantial component originating from Sr-$d$-orbitals indicating increased $s-d$ type hybridization with some participation of O-$s$ and H-$s$ atomic orbitals. 

\begin{table*}
\caption{Orbital composition analysis for the ground ${\rm X}^2\Sigma^+$ and lowest lying electronic ${\rm A}^2\Pi$ and ${\rm B}^2\Sigma^+$ states of the SrOH molecule. For each of the three molecular states columns correspond to amplitudes of the $s$, $p$, and $d$ valence orbitals of the strontium cation (Sr-{\it s}, Sr-{\it p}, Sr-{\it d}), the s valence orbital of  oxygen (O-{\it s}),  and the s orbital of hydrogen (H-{\it s}). }
\label{MO_composition}
\begin{center}
\begin{tabular}{l|ccccc}
\hline
State  & Sr-$s$ & Sr-$p$  & Sr-$d$  & O-$s$  &  H-$s$\\
\hline
\hline
X$^2\Sigma^+$  &0.9 & 0.1 & - & - &- \\
A$^2\Pi$  & - & 0.9 & 0.1 &- &-\\
B$^2\Sigma^+$ & $-0.2$&0.2 & 0.4 & 0.3 & 0.3\\
\end{tabular}
\end{center}
\end{table*}

\begin{figure*}
\includegraphics[width=0.5\textwidth,trim=0 0 0 0,clip]{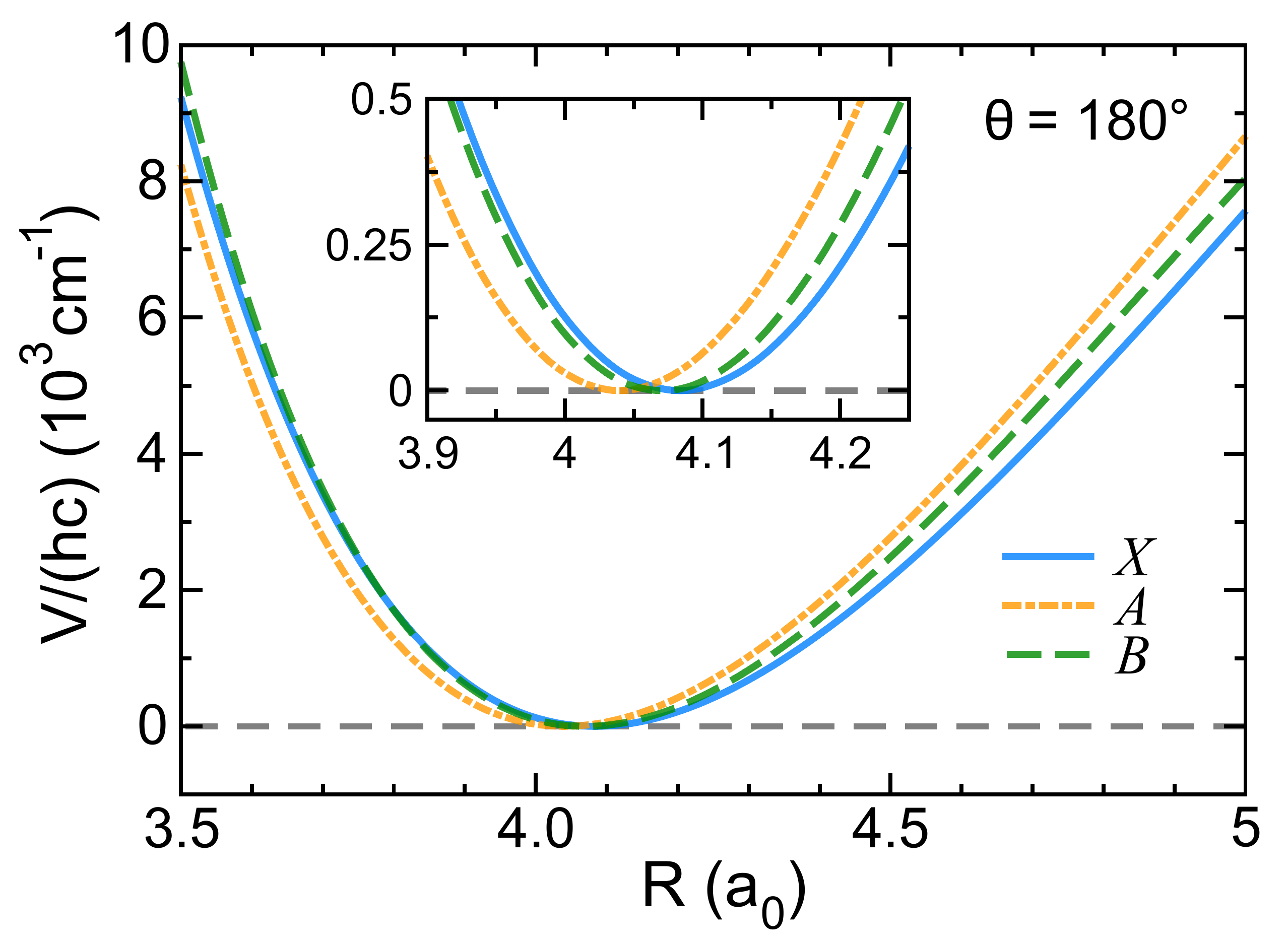}
\caption{Overlay of the ground and excited electronic potential energies $V$ as 
functions of separation R at angle $\theta=180^o$, linear geometry.  The inset shows a blowup of the potential minima. }
\label{fig:compmin180}
\end{figure*}

Our computation of vibrational wavefunctions and their overlap for 
electronic transitions is described in Methods.
We focus on vibrational levels near the global minima of the  
diabatic PESs at $\theta=180^\circ$ as shown in Fig.~\ref{fig:compmin180}. Effects from diabatic 
couplings, CIs, fine-structure interactions, and coriolis forces 
among the $A^\prime$ and $A^{\prime\prime}$ 
surfaces can then be omitted. The lowest vibrational and rotational 
states are uniquely labeled by the three vibrational quantum numbers 
$v_{\rm s}$, $v_{\rm b}$, and $v_{\rm OH}$, total trimer orbital 
angular momentum $J$, and parity $p$. Here, the vibrational quantum
numbers correspond to the Sr-O stretch, the Sr-O-H bend, and the O-H 
stretch, respectively. We use the abbreviated notation 
$(v_{\rm s},v_{\rm b},v_{\rm OH})$ and always have $v_{\rm OH}=0$ 
for frozen separation $r$. Our vibrational energies are discussed 
and compared to results from existing literature in Methods.

\begin{figure*}
\includegraphics[width=1 \textwidth,trim=0 0 0 0,clip]{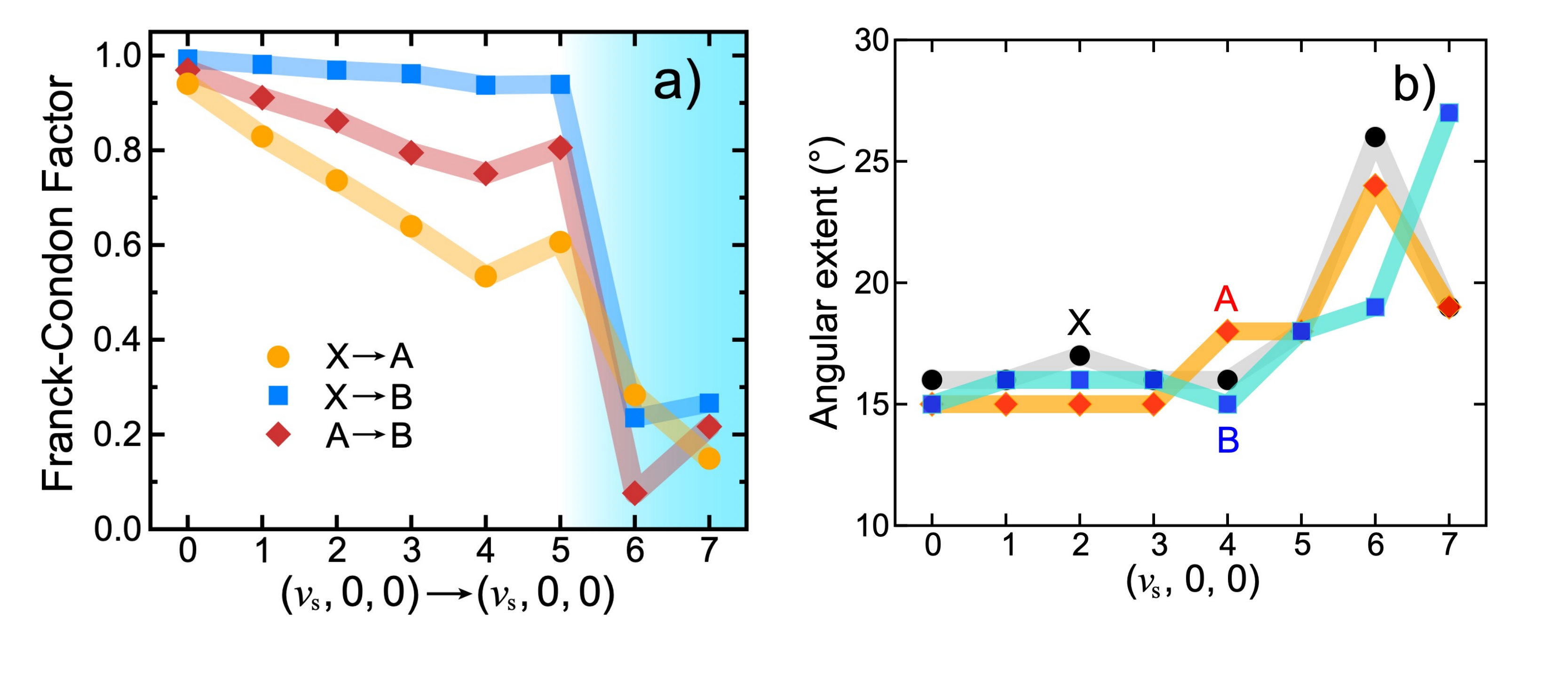}
\caption{Characterization of optical cycling transitions as functions of vibrational stretching mode quantum number $v_{\rm s}$.
a) Diagonal Frank-Condon factors of vibrational transitions among electronic states 
$1\,^2\!A^{\prime}({\rm X}^2\Sigma^+)$, 
$2\,^2\!A^{\prime}({\rm A}^2\Pi)$, and 
$3\,^2\!A^{\prime}({\rm B}^2\Sigma^+)$
as functions of vibrational states $(v'_{\rm s},v'_{\rm b},v'_{\rm OH})=(v_{\rm s},0,0)\to(v_{\rm s},0,0)$.
Transitions are color coded by the linear $C_{\infty v}$ symmetry, 
i.\,e.  blue, red, and orange curves and markers correspond
to the X$\to$B, A$\to$B, and X$\to$A transitions, respectively.
The blue shaded area indicates a sudden decrease of Frank-Condon factors for all 
three vibronic manifolds at $v_{\rm s}=6$ and 7.
b) The expected angular extent of the vibrational wave function relative to the angle $\theta=180^\circ$ equilibrium geometry as a function of the stretching mode quantum number. The larger extent for $v_{\rm s} > 5$ corresponds to a significant wave function probability away from linearity.
}
\label{fig:SrOH_DiagFCF}
\end{figure*}

\begin{figure*}
 \centering
    \includegraphics[width=0.7\textwidth,trim=0 -10 0 -10,clip] {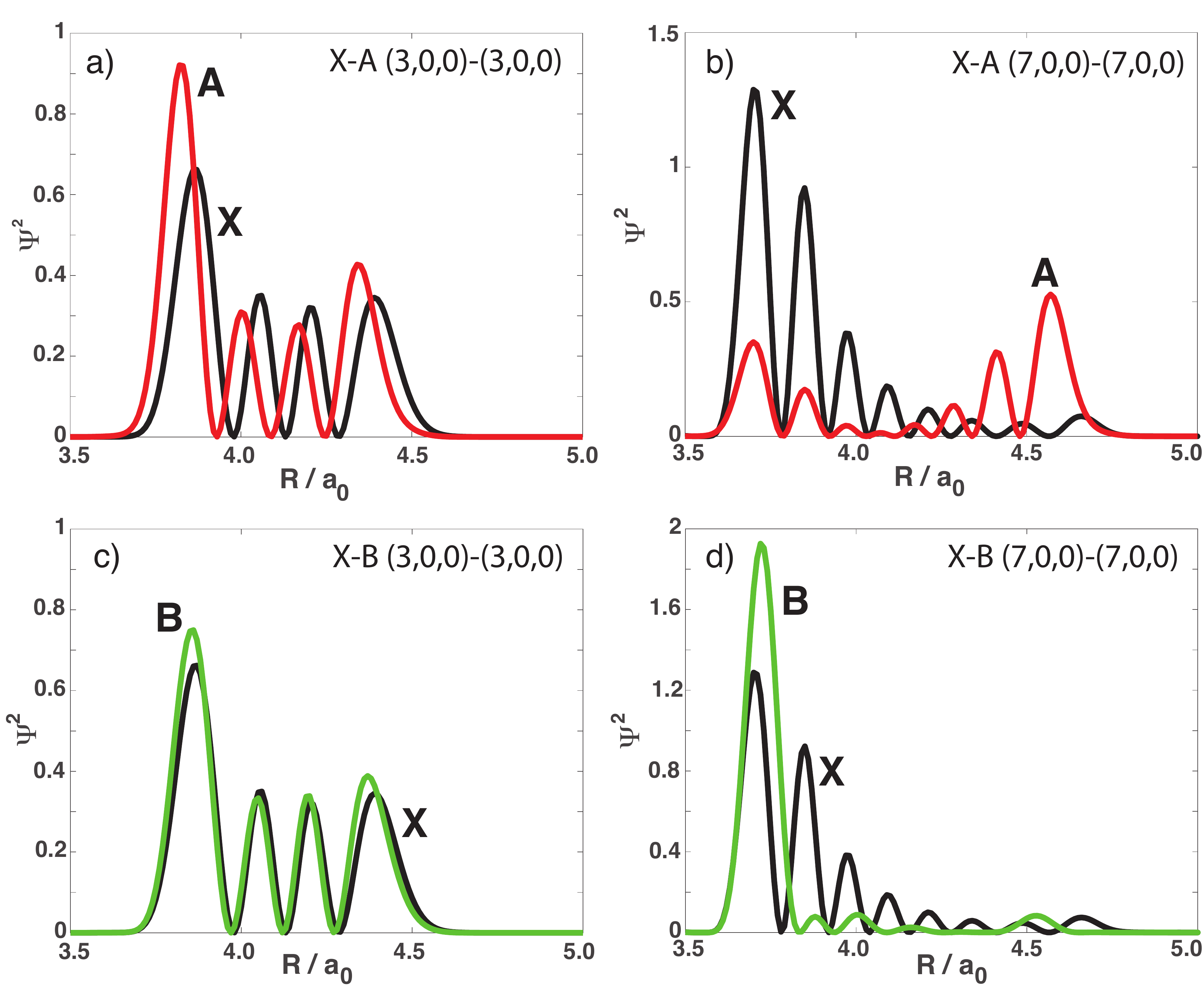}
 \caption{Square of the vibrational wave function for the  
 stretching modes $(v_{\rm s},0,0)=(3,0,0)$ and (7,0,0)  as  functions of separation $R$ at angle $\theta=180^\circ$
 for the ground $1\,^2\!A^{\prime}({\rm X}^2\Sigma^+)$ (black curves) and excited $2\,^2\!A^{\prime}({\rm A}^2\Pi)$, and 
 $3\,^2\!A^{\prime}({\rm B}^2\Sigma^+)$  (red and green curves) electronic states, respectively.
 a) Sr-O stretch with $(v_{\rm s},0,0)=(3,0,0)$ in the ${\rm X}^2\Sigma^+$ and ${\rm A}^2\Pi$ potentials; 
 b)  Sr-O stretch with $(v_{\rm s},0,0)=(7,0,0)$ in the ${\rm X}^2\Sigma^+$ and ${\rm A}^2\Pi$ potentials;
 c) Sr-O stretch with $(v_{\rm s},0,0)=(3,0,0)$ in the ${\rm X}^2\Sigma^+$ and ${\rm B}^2\Sigma^+$ potentials;
 d) Sr-O stretch with $(v_{\rm s},0,0)=(7,0,0)$ in the ${\rm X}^2\Sigma^+$ and ${\rm B}^2\Sigma^+$ potentials.
  }
 \label{fig:FCOreg}
\end{figure*}

\begin{table}[t]
\caption{ 
Franck-Condon factors between pairs of $(v_{\rm s},v_{\rm b},0)$
rovibrational SrOH states of 
${J=0}$ $2\,^2\!A^{\prime}({\rm A}^2\Pi)$,
${J=0}$ $3\,^2\!A^{\prime}({\rm B}^2\Sigma^+)$, and 
${J=1}$ $1\,^2\!A^{\prime}({\rm X}^2\Sigma^+)$.
In the table the three states are denoted by A, B and X, respectively.
Here parity $p=+1$ or $-1$ corresponds to whether $v_{\rm b}$ is even 
or odd.}
\label{tab:FCFmat}
\begin{center}
\def\arraystretch{1.2}
\begin{tabular}{l|cccccc}
\hline
A{\textbackslash}X  &   ~(000)~ & (100)~  &  (200)~ &   (300)~  &   (010)~  &  (110)~
\Tstrut\Bstrut\\
\hline
(000)  &   {\bf 0.940} &  {\bf 0.058}    &    &     &      &      \Tstrut\\
(100)  &   {\bf 0.056} &     {\bf 0.829} &     {\bf 0.112}  &    {\bf 0.003} &       &     \\
(200)  &   {\bf 0.003}  &    {\bf 0.104} &     {\bf 0.736} &     {\bf 0.151} &       &     \\
(300)  &    &     {\bf 0.007} &    {\bf 0.136}  &   {\bf  0.640}  &       &     \\
(010)  &    &     &       &     &     {\bf 0.941}  &    {\bf 0.057} \\
(110)   &    &      &      &       &  {\bf 0.055}   &     {\bf 0.832} \Bstrut\\
\hline
\hline
B{\textbackslash}X    &   ~(000)~ &   (100)~ &  (200)~ &   (300)~ &   (010)~ &   (110)~
\Tstrut\Bstrut\\
\hline
(000) &   {\bf 0.993}  &   {\bf 0.006}   &    &     &     &     \Tstrut\\
(100) &   {\bf 0.006}  &   {\bf 0.981}   &  {\bf 0.011}  &     &     &     \\
(200) &     &   {\bf 0.011}   &  {\bf 0.969}  &   {\bf 0.017}  &     &     \\
(300) &     &      &  {\bf 0.017}  &   {\bf 0.962}  &     &     \\
(010) &     &      &    &     &   {\bf 0.994}  &   {\bf 0.004}  \\
(110) &     &      &    &     &   {\bf 0.004}  &   {\bf 0.984}  \Bstrut\\
\hline
\hline
A{\textbackslash}B  &   ~(000)~ &   (100)~ &  (200)~ &   (300)~ &   (010)~ &   (110)~
\Tstrut\Bstrut\\
\hline
 (000) &    {\bf 0.969}  &   {\bf 0.030}  &         &          &          &          \Tstrut\\
 (100) &    {\bf 0.028}  &   {\bf 0.911}  &   {\bf 0.060} &          &          &          \\
 (200) &    {\bf 0.002}  &   {\bf 0.053}  &   {\bf 0.862} &    {\bf 0.081} &          &         \\
 (300) &           &   {\bf 0.004}  &   {\bf 0.067} &    {\bf 0.795} &          &          \\
 (010) &           &          &         &          &   {\bf 0.967} &  {\bf 0.033}   \\
 (110) &           &          &         &          &   {\bf 0.031} &   {\bf 0.902}  \Bstrut\\
\hline
\end{tabular}
\end{center}
\end{table}

\vspace*{6mm}

\noindent
{\bf Evaluation of Frank-Condon factors}. Figure~\ref{fig:SrOH_DiagFCF}a examines our diagonal FCFs of
stretching modes $(v'_{\rm s},v'_{\rm b},v'_{\rm OH})=(v_{\rm s},0,0)\to(v_{\rm s},0,0)$ among the 
${J=1}$ $1\,^2\!A^{\prime}({\rm X}^2\Sigma^+)$,
${J=0}$ $2\,^2\!A^{\prime}({\rm A}^2\Pi)$, and the 
${J=0}$ $3\,^2\!A^{\prime}({\rm B}^2\Sigma^+)$ states. 
For $v_{\rm s}\le 4$ the FCFs decrease linearly with increasing 
$v_{\rm s}$. In this regime, stretching modes with different 
bending quantum numbers are well separated in energy and FCFs 
based on the harmonic approximation of the potentials are in 
reasonable agreement with our more precise evaluation. 
The FCFs are largest for the 
$1\,^2\!A^{\prime}({\rm X}^2\Sigma^+)$ to 
$3\,^2\!A^{\prime}({\rm B}^2\Sigma^+)$ transition.

We observe a dramatic decrease of the FCFs for $v_{\rm s}=6$ and $7$,
emphasized by a blue shaded area in Fig.~\ref{fig:SrOH_DiagFCF}a. We explain this behavior in terms of the complex overlap of  vibrational wavefunctions in $R$ and $\theta$.
Figure~\ref{fig:SrOH_DiagFCF}b shows  the expected angular extents of vibrational wavefunctions relative to 
the $\theta=180^\circ$ equilibrium linear geometry as  functions of  $v_{\rm s}$. For  $v_{\rm b}=0$ states the vibrational wavefunctions
along $\theta$ are nodeless. We observe that  the bending range for $v_{\rm s}<4$ vibrational levels  
in all three electronic states is $\approx15^\circ$  and that for  $v_{\rm s}>5$ the wavefunction 
is more extended.  In fact, for  $v_{\rm s}>5$ the angular extent for the $1\,^2\!A^{\prime}({\rm X}^2\Sigma^+)$ and $2\,^2\!A^{\prime}({\rm A}^2\Pi)$
states remains close but differ significantly from those of the $3\,^2\!A^{\prime}({\rm B}^2\Sigma^+)$ state.
This irregularity in the $v_{\rm b}=0$ series coincides with the corresponding irregularity in FCFs, 
but clearly is not the whole story as we could conclude that the $1\,^2\!A^{\prime}({\rm X}^2\Sigma^+)$ to $2\,^2\!A^{\prime}({\rm A}^2\Pi)$
transitions remain diagonal.

Figure \ref{fig:FCOreg} compares  vibrational wavefunctions as  functions of stretching-direction $R$ for linear geometry $\theta=180^\circ$
for the $1\,^2\!A^{\prime}({\rm X}^2\Sigma^+)$ to $2\,^2\!A^{\prime}({\rm A}^2\Pi)$ and $1\,^2\!A^{\prime}({\rm X}^2\Sigma^+)$ to $3\,^2\!A^{\prime}({\rm B}^2\Sigma^+)$ transitions.
We observe that for the (3,0,0) vibrational level the wavefunction of the three electronic potentials have similar
shape, although the overlap for the $1\,^2\!A^{\prime}({\rm X}^2\Sigma^+)$ and $3\,^2\!A^{\prime}({\rm B}^2\Sigma^+)$ states
is significantly better. For the (7,0,0) level neither overlap is  good.
Combining our observations on the angular extend in Fig.~\ref{fig:SrOH_DiagFCF}b with those on the  stretching direction
then explains the dramatic decrease of the FCFs for $v_{\rm s}=6$ and $7$.

Table \ref{tab:FCFmat} presents our diagonal and off-diagonal FCF 
matrix elements. For simplicity FCFs smaller than $10^{-3}$ 
have been omitted. Overall the 
$1\,^2\!A^{\prime}({\rm X}^2\Sigma^+)$ to 
$3\,^2\!A^{\prime}({\rm B}^2\Sigma^+)$ vibronic transitions have 
higher diagonal FCFs and smaller off-diagonal FCFs than those for the
$1\,^2\!A^{\prime}({\rm X}^2\Sigma^+)$ to 
$2\,^2\!A^{\prime}({\rm A}^2\Pi)$ transition. 
The $(0,0,0)$ to $(0,0,0)$ FCFs for these electronic states was measured in Ref.~\cite{Nguyen:2018}.
They found 0.957(3) and 0.977(2) for the $1\,^2\!A^{\prime}({\rm X}^2\Sigma^+)$ to 
$2\,^2\!A^{\prime}({\rm A}^2\Pi)$ and $1\,^2\!A^{\prime}({\rm X}^2\Sigma^+)$ to 
$3\,^2\!A^{\prime}({\rm B}^2\Sigma^+)$ transition, respectively. These value are smaller than those found in our
calculations. Better agreement might be found by allowing the OH bond to change.
Extended tables of the FCF matrices are presented in Supplementary Note 1.

\vspace* {6mm}

\noindent
{\large\bf Discussion} \\
We have quantitatively analyzed the unique localized chemical bond 
between valence electrons of Sr and OH radical.
We have shown that one of the two Sr valence electrons remains 
localized on the Sr$^+$ cation and can be optically excited without
disturbing the atom-ligand bond leading to highly-diagonal FCFs and 
efficient optical-cycling conditions. The goal of our theoretical work has been to analyze the 
cooling process based on a rigorous calculation of the electronic structure of this important 
molecule previously totally unknown, as a means to look into the nature of optical cycling 
centers in polyatomic molecules and help extend its generality. Furthermore, our work indicates that 
there are several electronic and vibrational states that can be used for the laser cooling with FCFs close 
to unity. We also explain why the X-B transitions have more diagonal FCFs than X-A transitions. 

Our analysis of the electronic and vibrational energy landscape of 
SrOH also points to intriguing molecular features that require further
explorations. For instance, we located conical intersections in SrOH 
that can have important implications for non-adiabatic transitions 
between PESs, collisional dynamics within this molecule, as well as 
its fragmentation. Our characterization of the geometric Renner-Teller
effects provides better understanding of the coupling between 
electronic and vibrational motion and accounts for the possible 
nominally-forbidden loss channels.

Finally, Ref.~\cite{Kosyryev_2019} proposed that replacing the 
hydrogen atom in the monohydroxide with larger ligand should not significantly disturb the valence electron of metal-cation so that 
optical cycling and cooling even larger polyatomic molecule might 
remain possible. Detailed confirmation of this idea  also awaits 
future research.

\vspace*{6mm}

\noindent
{\large \bf Methods}\\
\noindent
{\bf Ab~initio methods for the SrOH potential energy surfaces.}\label{sec:shortrange}
We have used the coupled-cluster CCSD(T) method from CFOUR 
\cite{cfour} to calculate the ground and first excited 
diabatic PESs with total electron spin 1/2 of $A^{\prime}$ 
symmetry (ground state), $A^{\prime\prime}$ symmetry 
(first excited state), and the highly excited diabatic PES 
with spin 1/2 of $A^{\prime\prime}$ symmetry but with completely 
different reference configuration.
They are labeled $1\,^2\!A^{\prime}$, $1\,^2\!A^{\prime\prime}$ 
and $2\,^2\!A^{\prime\prime}$, respectively. The CCSD(T) calculations
are based on the atomic bases sets def2-QZVPP (8s8p5d3f)/[7s5p4d3f]
\cite{def2-QZVPP} paired with the Stuttgart ECP28MDF 
effective core potential \cite{Lim:2006} for Sr,  
the all electron aug-cc-pVQZ (13s7p4d3f2g)/[6s5p4d3f2g] for O, 
and aug-cc-pVQZ (7s4p3d2f)/[5s4p3d2f] for H \cite{dunning:89}. 

For all excited diabatic potentials, $1\,^2\!A^{\prime\prime}$, 
$2\,^2\!A^{\prime}$, $3\,^2\!A^{\prime}$, $2\,^2\!A^{\prime\prime}$, 
and $4\,^2\!A^{\prime}$, we employed the equation-of-motion 
coupled-cluster (EOM-CCSD(dT)) approach, implemented in Q-CHEM, 
version 5.0 \cite{QChem}. The excitation-energy root space for the 
EOM-CCSD(dT) method spans four roots of $A^{\prime}$ symmetry and 
two roots of the $A^{\prime\prime}$ symmetry. 
We use the Stuttgart ECP28MDF relativistic effective core potential
\cite{Lim:2006} along with Peterson's pseudopotential based, 
polarized valence, correlation consistent triple-zeta (aug-cc-pVTZ-PP) 
basis for the Sr atom~\cite{Li:2013} and corresponding Dunning's 
aug-cc-pVTZ basis for the O and H atoms~\cite{dunning:89}.
These calculations allowed us to obtain the relative splittings 
between $A^{\prime}$ and $A^{\prime\prime}$ PESs and between the 
$2\,^2\!A^{\prime}$ and $3\,^2\!A^{\prime}$ PESs.

The CC based methods used in construction of the $^2\!A^{\prime}$ and 
$^2\!A^{\prime\prime}$ potentials are capable of providing highly 
accurate quantitative data about PESs. Unfortunately, these methods 
are unsuitable for the calculation of PESs near their crossing or 
avoided crossing. The way out is a MRCI calculation, 
which realistically, can only be done with an active space of limited
size.

We characterize non-adiabatic couplings between states by using the 
MRCI calculations with single and double excitations,
implemented in MOLPRO~\cite{molpro}.
The ECP28MDF K{\"o}ln pseudo-potential
\cite{Lim:2006} with the augmented correlation-consistent 
triple-zeta basis aug-cc-pVTZ-PP \cite{Li:2013} for Sr and the 
all-electron aug-cc-pVQZ basis for O and H \cite{dunning:89} 
are applied. The calculations are performed with reference orbitals
constructed from  state-averaged multi-reference self-consistent-field
calculations. Finally, all electronic structure calculations were 
performed with assumption that the OH separation is fixed at 
$r=1.832 a_0$.

\vspace*{6mm}
\noindent
{\bf Interpolation of potential energy surfaces.} 
 The four $^2\!A^\prime$ and two $^2\!A^{\prime\prime}$ two-dimensional 
{\em diabatic} SrOH potentials have been computed on a grid in Jacobi coordinates 
$R$ and $\theta$. In order to find the potential energies at other
coordinates we  expand the angular dependence of each {\em diabatic} 
PESs in terms of Legendre polynomials $P_l(\cos\theta)$ via
\begin{equation}
 V_{\rm mr}(R,\theta)=\sum_{l=0}^{L}U_{l}(R)P_{l}(\cos\theta) 
\end{equation}
for $R<16.5a_0$  with $L=9$ and radial expansion coefficients 
$U_{l}(R)$. For separations $R>16.5a_0$ the PESs are fit to 
\begin{equation}
 V_{\rm lr}(R,\theta) = \sum_{n=6}^{9}\sum_{l=0}^{n-4} \frac{C_{nl}}{R^{n}}P_{l}(\cos\theta) 
\end{equation}
with dispersion coefficients $C_{nl}$. The radial coefficients 
$U_{l}(R)$ and  $C_{nl}$ are determined by the Reproducing Kernel 
Hilbert Space (RKHS) interpolation method~\cite{ho:96}. 
The two expansions are  smoothly joined using
\begin{equation}
 V(R,\theta)=[1-f(R)]\,V_{\rm mr}(R,\theta)+f(R)\,V_{\rm lr}(R,\theta) 
\end{equation}
and
$
 f(R)=(1+\tanh\left[\alpha\left(R-R_{\rm sw}\right)\right])/2.
$
The parameter values for describing $U_{l}(R)$, $C_{nl}$, $\alpha$, 
and $R_{\rm sw}$ are available upon request.

\vspace*{6mm}
\noindent
{\bf Bound state energies.}
For each diabatic PESs the corresponding vibrational Hamiltonian
includes the kinetic-energy operators for Jacobi coordinates 
{\bf R} and {\bf r} with masses 
${\mu=m_{\rm Sr}m_{\rm OH}/({m_{\rm Sr}+m_{\rm OH}})}$ and $m_{\rm OH}$,
respectively. Here, $m_{\rm Sr}$  is the mass of the Sr atom and  
$m_{\rm OH}$ that of the OH molecule. The vibrational Hamiltonian 
commutes with the trimer orbital-angular-momentum operator 
{\bf J} = ${\bf j + l}$ as well as the parity operator $p$, 
the symmetry under spatial inversion of all electrons and nuclei 
around the center-of-mass position of the trimer. Here, $\bf j$
and $\bf l$ are the orbital-angular-momentum operators of 
ground-state OH and of Sr relative to the center of mass of OH, 
respectively.

\begin{table}
\caption{
Comparison of energies for three $(v_{\rm s}, v_{\rm b},0)$ stretching 
and bending levels with total angular momentum ${J=0}$ and parity $ +1$ 
for the $1\,^2\!A^\prime({\rm X}^2\Sigma^+)$, 
$1\,^2\!A^{\prime\prime}({\rm A}^2\Pi)$, and 
$3\,^2\!A^\prime({\rm B}^2\Sigma^+)$ states of the 
$^{88}$Sr-$^{16}$O$^{1}$H isotopologue using our diabatic
and adiabatic potential energy surfaces, and data from the literature.
Energies are relative to that of the  $(0,0,0)$ mode and in units 
of cm$^{-1}$. 
}
\label{tb:rovib}
\begin{threeparttable}
\begin{tabular}[t]{>{\centering\arraybackslash}p{0.15\columnwidth}
                   *{3}{>{\centering\arraybackslash}p{0.08\columnwidth}}}
\hline
\hline
        Method          &       $(1,0,0)$       &       $(0,1,0)$       &     $(0,2,0)$ \Tstrut\Bstrut\\
\hline    
\multicolumn{4}{c}{$1\,^2\!A^\prime({\rm X}^2\Sigma^+)$}   \Tstrut\Bstrut\\
\hline
         CCSD(T)  &         524           &         382           &        748    \Tstrut\\
        2D-DVR~\cite{Nguyen:2018}      &          -            &         322           &        638           \\
        Exp.~\cite{Presunka:1993}      &         527           &         364           &        703    \\
\hline
\multicolumn{4}{c}{$1\,^2\!A^{\prime\prime}({\rm A}^2\Pi)$   }   \Tstrut\Bstrut\\
\hline
         CCSD(T)    &         544           &         412           &        808    \Tstrut\\
        2D-DVR~\cite{Nguyen:2018}      &          -            &         308           &        614           \\
        Exp.~\cite{Presunka:1993}      &         543           &          -            &         -            \\
        Exp.~\cite{Presunka:1995}      &          -            &         378           &         -     \\
\hline
\multicolumn{4}{c}{$3\,^2\!A^\prime({\rm B}^2\Sigma^+)$}   \Tstrut\Bstrut\\
\hline
     EOM-CCSD(dT)   &         549           &         417           &        799    \Tstrut\\
        2D-DVR~\cite{Nguyen:2018}     &          -            &         358           &        699           \\
        Exp.~\cite{Oberlander:1996}     &         536           &          -            &         -            \\
        Exp.~\cite{Presunka:1993}      &          -            &         401           &       771 \\
\hline\hline
\end{tabular}
\end{threeparttable}
\end{table}

We diagonalize the vibrational Hamiltonian by first creating a 
one-dimensional basis of bound states of the radial Hamiltonian 
$-\hbar^2/(2\mu)d^2/ dR^2+V(R)$  using the distributed Gaussian basis 
approach and where $V(R)$ is the interpolated PES computed at  
${\theta=180^{\circ}}$ and $r=1.832a_0$ ($\hbar$ is the reduced 
Planck constant.) Then the vibrational Hamiltonian is diagonalized 
in the product basis of radial bound states and superpositions of 
products of spherical harmonic functions for the orientation of 
{\bf R} and {\bf r}, such that the basis states are eigenstates of 
${\bf J}^2$, $J_z$, and parity $p$.

Within these approximations we have computed several of the 
energetically-lowest vibrational levels with total angular momentum  
${J^p=0^+}$ and $1^{\pm}$ of the diabatic 
$1\,^2\!A^\prime({\rm X}^2\Sigma^+)$, 
$2\,^2\!A^\prime({\rm A}^2\Pi)$, 
$1\,^2\!A^{\prime\prime}({\rm A}^2\Pi)$, and 
$3\,^2\!A^\prime({\rm B}^2\Sigma^+)$. 
We focus on these rotational states as ${{J=1}\to {J'=0}}$ transitions 
from the $1\,^2\!A^\prime({\rm X}^2\Sigma^+)$ ground state are used 
in laser cooling of SrOH~\cite{Doyle2017}. 

In Table~\ref{tb:rovib} we compare some of our calculated energies 
with total angular momentum $J=0$ to other semi-empirical results 
as well as experimental data. 
Our results for the Sr-O stretch and Sr-O-H bend energies agree well 
with experimental data, even though the OH bond is not allowed to 
stretch and diabatic couplings are omitted. The energy difference 
between the $(v_{\rm s},v_{\rm b},0)=(1,0,0)$ and $(0,0,0)$ states,
the Sr-O stretching mode, are reproduced to better than 1\% for the 
$1\,^2\!A^\prime({\rm X}^2\Sigma^+)$ and 
$1\,^2\!A^{\prime\prime}({\rm A}^2\Pi)$ states. 
The discrepancy increases to 2.5\% for the 
$3\,^2\!A^\prime({\rm B}^2\Sigma^+)$ state. 

For the bending mode with quanta $v_{\rm b}$ our theoretical energy
differences are larger than those found experimentally, 
although the difference is no more than 10\% for the three 
electronic states. Adding a second quanta in the bending mode shows 
that anharmonic contributions are non-negligible. 
We also corroborate the DFT and 2D-DVR results of Nguyen {\em et al.}
\cite{Nguyen:2018}. Nguyen's DFT results are based on the harmonic
approximation.

The vibrational wavefunctions are used the compute FCFs 
among various pairs of ionic states. Their values and trends were 
presented in subsection ``Wavefunction overlaps''. Additional bound-state energies and
wavefunctions are shown in Supplementary Figures 1-3.

We can also quantify the quality of our potentials by a comparison of the dissociation energy and permanent
dipole moments. The  dissociation energy of the 
(0,0,0) vibrational state of our $1\,^2\!A^\prime({\rm X}^2\Sigma^+)$ potential
is $D_0/hc=33.3\times 10^3$ cm$^{-1}$ and compares favorably with the experimental
measurements ranging from $32.3\times 10^3$ cm$^{-1}$ to $36.0\times 10^3$ cm$^{-1}$
\cite{SrOH_2,SrOH_3,SrOH_5}.
Our determination of the permanent dipole moments of the $1\,^2\!A^\prime({\rm X}^2\Sigma^+)$, 
$2\,^2\!A^\prime({\rm A}^2\Pi)$, and $3\,^2\!A^\prime({\rm B}^2\Sigma^+)$ states at their equilibrium geometries are
1.57 Debye, 0.314 Debye, and 0.396 Debye, respectively.  Reference~\cite{Steimle1992} measured 
1.900(14) Debye and 0.396(61) Debye for the (0,0,0) vibrational state of the $1\,^2\!A^\prime({\rm X}^2\Sigma^+)$ and $3\,^2\!A^\prime({\rm B}^2\Sigma^+)$ potentials. The agreement is mixed.
A direct comparison for the $2\,^2\!A^\prime({\rm A}^2\Pi)$ state can not be made as
Ref.~\cite{Steimle1992} presented a fine-structure resolved measurement, i.e. they found
permanent dipole moments of 0.590(45) Debye and 0.424(5) Debye for $\Omega=1/2$ and 3/2, respectively, suggesting that our value is too small. (The numbers in parenthesis are one-standard-deviation uncertainties in the last digits.)

\vspace*{6mm}
\noindent
{\large \bf Data Availability}\\
The data sets generated during the current study are partially included in this published article (and its Supplementary Note 1) and also available from the corresponding author on reasonable request.

\vspace*{6mm}
\noindent
{\large \bf Acknowledgment}\\
Work at  Temple University is supported by the Army Research Office
Grant No. W911NF-17-1-0563, the U.S. Air Force Office of Scientific 
Research Grant No. FA9550-14-1-0321 and the NSF Grant No. PHY-1908634. 

\vspace*{6mm}
\noindent
{\large \bf Authors Contributions} \\
S. K. conceived, designed, coordinated the work, and wrote the manuscript. Electronic structure calculations were performed by J. K., M. L. and A. P. Fitting of the potentials were performed by J. K and M. L. Bound states, wavefunctions and Franck-Condon factors calculations were performed by J. K. All authors contributed to the interpretation of the data and discussed the results.

\vspace*{6mm}
\noindent
{\large \bf Competing Interests}\\
The authors declare no competing interests.


\bibliography{SrOH_library}

\begin{thebibliography}{50}%
\makeatletter
\providecommand \@ifxundefined [1]{%
 \@ifx{#1\undefined}
}%
\providecommand \@ifnum [1]{%
 \ifnum #1\expandafter \@firstoftwo
 \else \expandafter \@secondoftwo
 \fi
}%
\providecommand \@ifx [1]{%
 \ifx #1\expandafter \@firstoftwo
 \else \expandafter \@secondoftwo
 \fi
}%
\providecommand \natexlab [1]{#1}%
\providecommand \enquote  [1]{``#1''}%
\providecommand \bibnamefont  [1]{#1}%
\providecommand \bibfnamefont [1]{#1}%
\providecommand \citenamefont [1]{#1}%
\providecommand \href@noop [0]{\@secondoftwo}%
\providecommand \href [0]{\begingroup \@sanitize@url \@href}%
\providecommand \@href[1]{\@@startlink{#1}\@@href}%
\providecommand \@@href[1]{\endgroup#1\@@endlink}%
\providecommand \@sanitize@url [0]{\catcode `\\12\catcode `\$12\catcode
  `\&12\catcode `\#12\catcode `\^12\catcode `\_12\catcode `\%12\relax}%
\providecommand \@@startlink[1]{}%
\providecommand \@@endlink[0]{}%
\providecommand \url  [0]{\begingroup\@sanitize@url \@url }%
\providecommand \@url [1]{\endgroup\@href {#1}{\urlprefix }}%
\providecommand \urlprefix  [0]{URL }%
\providecommand \Eprint [0]{\href }%
\providecommand \doibase [0]{http://dx.doi.org/}%
\providecommand \selectlanguage [0]{\@gobble}%
\providecommand \bibinfo  [0]{\@secondoftwo}%
\providecommand \bibfield  [0]{\@secondoftwo}%
\providecommand \translation [1]{[#1]}%
\providecommand \BibitemOpen [0]{}%
\providecommand \bibitemStop [0]{}%
\providecommand \bibitemNoStop [0]{.\EOS\space}%
\providecommand \EOS [0]{\spacefactor3000\relax}%
\providecommand \BibitemShut  [1]{\csname bibitem#1\endcsname}%
\let\auto@bib@innerbib\@empty
\bibitem [{\citenamefont {Wieman}\ \emph {et~al.}(1999)\citenamefont {Wieman},
  \citenamefont {Pritchard},\ and\ \citenamefont {Wineland}}]{Wieman1999}%
  \BibitemOpen
  \bibfield  {author} {\bibinfo {author} {\bibfnamefont {C.~E.}\ \bibnamefont
  {Wieman}}, \bibinfo {author} {\bibfnamefont {D.~E.}\ \bibnamefont
  {Pritchard}}, \ and\ \bibinfo {author} {\bibfnamefont {D.~J.}\ \bibnamefont
  {Wineland}},\ }\enquote {\bibinfo {title} {Atom cooling, trapping, and
  quantum manipulation},}\ in\ \href@noop {} {\emph {\bibinfo {booktitle} {More
  Things in Heaven and Earth: A Celebration of Physics at the Millennium}}},\
  \bibinfo {editor} {edited by\ \bibinfo {editor} {\bibfnamefont
  {B.}~\bibnamefont {Bederson}}}\ (\bibinfo  {publisher} {Springer New York},\
  \bibinfo {address} {New York, NY},\ \bibinfo {year} {1999})\ pp.\ \bibinfo
  {pages} {426--441}\BibitemShut {NoStop}%
\bibitem [{\citenamefont {Presunka}\ and\ \citenamefont
  {Coxon}(1995)}]{Presunka:1995}%
  \BibitemOpen
  \bibfield  {author} {\bibinfo {author} {\bibfnamefont {P.~I.}\ \bibnamefont
  {Presunka}}\ and\ \bibinfo {author} {\bibfnamefont {J.~A.}\ \bibnamefont
  {Coxon}},\ }\bibfield  {title} {\enquote {\bibinfo {title} {Laser excitation
  and dispersed fluorescence investigations of the
  {A}$^2\ensuremath{\Pi}$-{X}$^2\ensuremath{\Sigma}^+$ system of {SrOH}},}\
  }\href@noop {} {\bibfield  {journal} {\bibinfo  {journal} {Chem. Phys.}\
  }\textbf {\bibinfo {volume} {190}},\ \bibinfo {pages} {97--111} (\bibinfo
  {year} {1995})}\BibitemShut {NoStop}%
\bibitem [{\citenamefont {Wormsbecher}\ \emph
  {et~al.}(1983{\natexlab{a}})\citenamefont {Wormsbecher}, \citenamefont
  {Penn},\ and\ \citenamefont {Harris}}]{WORMSBECHER1}%
  \BibitemOpen
  \bibfield  {author} {\bibinfo {author} {\bibfnamefont {R.~F.}\ \bibnamefont
  {Wormsbecher}}, \bibinfo {author} {\bibfnamefont {R.~E.}\ \bibnamefont
  {Penn}}, \ and\ \bibinfo {author} {\bibfnamefont {D.~O.}\ \bibnamefont
  {Harris}},\ }\bibfield  {title} {\enquote {\bibinfo {title} {High-resolution
  laser spectroscopy of {CaNH}$_2$: {A}nalysis of the
  {C}$\,^2${A}$_1$-{X}$\,^2${A}$_1$ system},}\ }\href@noop {} {\bibfield
  {journal} {\bibinfo  {journal} {J. Mol. Spec.}\ }\textbf {\bibinfo {volume}
  {97}},\ \bibinfo {pages} {65--72} (\bibinfo {year}
  {1983}{\natexlab{a}})}\BibitemShut {NoStop}%
\bibitem [{\citenamefont {Wormsbecher}\ \emph
  {et~al.}(1983{\natexlab{b}})\citenamefont {Wormsbecher}, \citenamefont
  {Trkula}, \citenamefont {Martner}, \citenamefont {Penn},\ and\ \citenamefont
  {Harris}}]{WORMSBECHER2}%
  \BibitemOpen
  \bibfield  {author} {\bibinfo {author} {\bibfnamefont {R.~F.}\ \bibnamefont
  {Wormsbecher}}, \bibinfo {author} {\bibfnamefont {M.}~\bibnamefont {Trkula}},
  \bibinfo {author} {\bibfnamefont {C.}~\bibnamefont {Martner}}, \bibinfo
  {author} {\bibfnamefont {R.~E.}\ \bibnamefont {Penn}}, \ and\ \bibinfo
  {author} {\bibfnamefont {D.~O.}\ \bibnamefont {Harris}},\ }\bibfield  {title}
  {\enquote {\bibinfo {title} {Chemiluminescent reactions of alkaline-earth
  metals with water and hydrazine},}\ }\href@noop {} {\bibfield  {journal}
  {\bibinfo  {journal} {J. Mol. Spec.}\ }\textbf {\bibinfo {volume} {97}},\
  \bibinfo {pages} {29--36} (\bibinfo {year} {1983}{\natexlab{b}})}\BibitemShut
  {NoStop}%
\bibitem [{\citenamefont {Hilborn}\ \emph {et~al.}(1983)\citenamefont
  {Hilborn}, \citenamefont {Qingshi},\ and\ \citenamefont
  {Harris}}]{Hilborn1983}%
  \BibitemOpen
  \bibfield  {author} {\bibinfo {author} {\bibfnamefont {R.~C.}\ \bibnamefont
  {Hilborn}}, \bibinfo {author} {\bibfnamefont {Z.}~\bibnamefont {Qingshi}}, \
  and\ \bibinfo {author} {\bibfnamefont {D.~O.}\ \bibnamefont {Harris}},\
  }\bibfield  {title} {\enquote {\bibinfo {title} {Laser spectroscopy of the
  {A-X} transitions of {CaOH} and {CaOD}},}\ }\href@noop {} {\bibfield
  {journal} {\bibinfo  {journal} {J. Mol. Spec.}\ }\textbf {\bibinfo {volume}
  {97}},\ \bibinfo {pages} {73--91} (\bibinfo {year} {1983})}\BibitemShut
  {NoStop}%
\bibitem [{\citenamefont {Bernath}\ and\ \citenamefont
  {Kinsey-Nielsen}(1984)}]{BERNATH1984}%
  \BibitemOpen
  \bibfield  {author} {\bibinfo {author} {\bibfnamefont {P.F.}\ \bibnamefont
  {Bernath}}\ and\ \bibinfo {author} {\bibfnamefont {S.}~\bibnamefont
  {Kinsey-Nielsen}},\ }\bibfield  {title} {\enquote {\bibinfo {title} {Dye
  laser spectroscopy of the
  {B}$^2\ensuremath{\Sigma}^+$-{X}$^2\ensuremath{\Sigma}^+$ transition of
  {CaOH}},}\ }\href@noop {} {\bibfield  {journal} {\bibinfo  {journal} {Chem.
  Phys. Lett.}\ }\textbf {\bibinfo {volume} {105}},\ \bibinfo {pages}
  {663--666} (\bibinfo {year} {1984})}\BibitemShut {NoStop}%
\bibitem [{\citenamefont {Brazier}\ and\ \citenamefont
  {Bernath}(1987)}]{Brazier1987}%
  \BibitemOpen
  \bibfield  {author} {\bibinfo {author} {\bibfnamefont {C.~R.}\ \bibnamefont
  {Brazier}}\ and\ \bibinfo {author} {\bibfnamefont {P.~F.}\ \bibnamefont
  {Bernath}},\ }\bibfield  {title} {\enquote {\bibinfo {title} {Observation of
  gas phase organometallic free radicals: {M}onomethyl derivatives of calcium
  and strontium},}\ }\href@noop {} {\bibfield  {journal} {\bibinfo  {journal}
  {J. Chem. Phys.}\ }\textbf {\bibinfo {volume} {86}},\ \bibinfo {pages}
  {5918--5922} (\bibinfo {year} {1987})}\BibitemShut {NoStop}%
\bibitem [{\citenamefont {Bopegedera}\ \emph {et~al.}(1987)\citenamefont
  {Bopegedera}, \citenamefont {Brazier},\ and\ \citenamefont
  {Bernath}}]{Bopegedera1987}%
  \BibitemOpen
  \bibfield  {author} {\bibinfo {author} {\bibfnamefont {A.~M. R.~P.}\
  \bibnamefont {Bopegedera}}, \bibinfo {author} {\bibfnamefont {C.~R.}\
  \bibnamefont {Brazier}}, \ and\ \bibinfo {author} {\bibfnamefont {P.~F.}\
  \bibnamefont {Bernath}},\ }\bibfield  {title} {\enquote {\bibinfo {title}
  {Laser spectroscopy of strontium and calcium monoalkylamides},}\ }\href@noop
  {} {\bibfield  {journal} {\bibinfo  {journal} {J. Phys. Chem.}\ }\textbf
  {\bibinfo {volume} {91}},\ \bibinfo {pages} {2779--2781} (\bibinfo {year}
  {1987})}\BibitemShut {NoStop}%
\bibitem [{\citenamefont {Brazier}\ and\ \citenamefont
  {Bernath}(1989)}]{Brazier1989}%
  \BibitemOpen
  \bibfield  {author} {\bibinfo {author} {\bibfnamefont {C.~R.}\ \bibnamefont
  {Brazier}}\ and\ \bibinfo {author} {\bibfnamefont {P.~F.}\ \bibnamefont
  {Bernath}},\ }\bibfield  {title} {\enquote {\bibinfo {title} {The
  {A}$^2\ensuremath{\Pi}$-x$^2\ensuremath{\Sigma}^+$ transition of monomethyl
  calcium: {A} rotational analysis},}\ }\href@noop {} {\bibfield  {journal}
  {\bibinfo  {journal} {J. Chem. Phys.}\ }\textbf {\bibinfo {volume} {91}},\
  \bibinfo {pages} {4548--4554} (\bibinfo {year} {1989})}\BibitemShut {NoStop}%
\bibitem [{\citenamefont {Bernath}(1991)}]{BERNATH1991}%
  \BibitemOpen
  \bibfield  {author} {\bibinfo {author} {\bibfnamefont {P.~F.}\ \bibnamefont
  {Bernath}},\ }\bibfield  {title} {\enquote {\bibinfo {title} {Gas-phase
  inorganic chemistry: monovalent derivatives of calcium and strontium},}\
  }\href@noop {} {\bibfield  {journal} {\bibinfo  {journal} {Science}\ }\textbf
  {\bibinfo {volume} {254}},\ \bibinfo {pages} {665--670} (\bibinfo {year}
  {1991})}\BibitemShut {NoStop}%
\bibitem [{\citenamefont {Nakagawa}\ \emph {et~al.}(1983)\citenamefont
  {Nakagawa}, \citenamefont {Wormsbecher},\ and\ \citenamefont
  {Harris}}]{Nakagawa1983}%
  \BibitemOpen
  \bibfield  {author} {\bibinfo {author} {\bibfnamefont {J.}~\bibnamefont
  {Nakagawa}}, \bibinfo {author} {\bibfnamefont {R.~F.}\ \bibnamefont
  {Wormsbecher}}, \ and\ \bibinfo {author} {\bibfnamefont {D.~O.}\ \bibnamefont
  {Harris}},\ }\bibfield  {title} {\enquote {\bibinfo {title} {High-resolution
  laser excitation spectra of linear triatomic molecules: {A}nalysis of the
  {B}$^2{\Sigma}^+$-{X}$^2{\Sigma}^+$ system of {S}r{OH} and {S}r{OD}},}\
  }\href@noop {} {\bibfield  {journal} {\bibinfo  {journal} {J. Mol. Spec.}\
  }\textbf {\bibinfo {volume} {97}},\ \bibinfo {pages} {37--64} (\bibinfo
  {year} {1983})}\BibitemShut {NoStop}%
\bibitem [{\citenamefont {Jr.}\ and\ \citenamefont
  {Langhoff}(1998)}]{Langhoff1998}%
  \BibitemOpen
  \bibfield  {author} {\bibinfo {author} {\bibfnamefont {C.~W.~Bauschlicher}\
  \bibnamefont {Jr.}}\ and\ \bibinfo {author} {\bibfnamefont {S.~R.}\
  \bibnamefont {Langhoff}},\ }\bibfield  {title} {\enquote {\bibinfo {title}
  {{\em Ab initio} study of the alkali and alkaline-earth monohydroxides},}\
  }\href@noop {} {\bibfield  {journal} {\bibinfo  {journal} {J. Chem. Phys.}\
  }\textbf {\bibinfo {volume} {84}},\ \bibinfo {pages} {901--909} (\bibinfo
  {year} {1998})}\BibitemShut {NoStop}%
\bibitem [{\citenamefont {Dalibard}\ and\ \citenamefont
  {Cohen-Tannoudji}(1985)}]{Doppler1985}%
  \BibitemOpen
  \bibfield  {author} {\bibinfo {author} {\bibfnamefont {J.}~\bibnamefont
  {Dalibard}}\ and\ \bibinfo {author} {\bibfnamefont {C.}~\bibnamefont
  {Cohen-Tannoudji}},\ }\bibfield  {title} {\enquote {\bibinfo {title}
  {Dressed-atom approach to atomic motion in laser light: the dipole force
  revisited},}\ }\href@noop {} {\bibfield  {journal} {\bibinfo  {journal} {J.
  Opt. Soc. Am B}\ }\textbf {\bibinfo {volume} {2}},\ \bibinfo {pages}
  {1707--1720} (\bibinfo {year} {1985})}\BibitemShut {NoStop}%
\bibitem [{\citenamefont {Norrgard}\ \emph {et~al.}(2016)\citenamefont
  {Norrgard}, \citenamefont {McCarron}, \citenamefont {Steinecker},
  \citenamefont {Tarbutt},\ and\ \citenamefont {DeMille}}]{DeMille2016}%
  \BibitemOpen
  \bibfield  {author} {\bibinfo {author} {\bibfnamefont {E.~B.}\ \bibnamefont
  {Norrgard}}, \bibinfo {author} {\bibfnamefont {D.~J.}\ \bibnamefont
  {McCarron}}, \bibinfo {author} {\bibfnamefont {M.~H.}\ \bibnamefont
  {Steinecker}}, \bibinfo {author} {\bibfnamefont {M.~R.}\ \bibnamefont
  {Tarbutt}}, \ and\ \bibinfo {author} {\bibfnamefont {D.}~\bibnamefont
  {DeMille}},\ }\bibfield  {title} {\enquote {\bibinfo {title} {Submillikelvin
  dipolar molecules in a radio-frequency magneto-optical trap},}\ }\href@noop
  {} {\bibfield  {journal} {\bibinfo  {journal} {Phys. Rev. Lett.}\ }\textbf
  {\bibinfo {volume} {116}},\ \bibinfo {pages} {063004} (\bibinfo {year}
  {2016})}\BibitemShut {NoStop}%
\bibitem [{\citenamefont {Anderegg}\ \emph {et~al.}(2017)\citenamefont
  {Anderegg}, \citenamefont {Augenbraun}, \citenamefont {Chae}, \citenamefont
  {Hemmerling}, \citenamefont {Hutzler}, \citenamefont {Ravi}, \citenamefont
  {Collopy}, \citenamefont {Ye}, \citenamefont {Ketterle},\ and\ \citenamefont
  {Doyle}}]{Doyle2}%
  \BibitemOpen
  \bibfield  {author} {\bibinfo {author} {\bibfnamefont {L.}~\bibnamefont
  {Anderegg}}, \bibinfo {author} {\bibfnamefont {B.~L.}\ \bibnamefont
  {Augenbraun}}, \bibinfo {author} {\bibfnamefont {E.}~\bibnamefont {Chae}},
  \bibinfo {author} {\bibfnamefont {B.}~\bibnamefont {Hemmerling}}, \bibinfo
  {author} {\bibfnamefont {N.~R.}\ \bibnamefont {Hutzler}}, \bibinfo {author}
  {\bibfnamefont {A.}~\bibnamefont {Ravi}}, \bibinfo {author} {\bibfnamefont
  {A.}~\bibnamefont {Collopy}}, \bibinfo {author} {\bibfnamefont
  {J.}~\bibnamefont {Ye}}, \bibinfo {author} {\bibfnamefont {W.}~\bibnamefont
  {Ketterle}}, \ and\ \bibinfo {author} {\bibfnamefont {J.~M.}\ \bibnamefont
  {Doyle}},\ }\bibfield  {title} {\enquote {\bibinfo {title} {Radio frequency
  magneto-optical trapping of {CaF} with high density},}\ }\href@noop {}
  {\bibfield  {journal} {\bibinfo  {journal} {Phys. Rev. Lett.}\ }\textbf
  {\bibinfo {volume} {119}},\ \bibinfo {pages} {103201} (\bibinfo {year}
  {2017})}\BibitemShut {NoStop}%
\bibitem [{\citenamefont {Truppe}\ \emph {et~al.}(2017)\citenamefont {Truppe},
  \citenamefont {Williams}, \citenamefont {Hambach}, \citenamefont {Caldwell},
  \citenamefont {Fitch}, \citenamefont {Hinds}, \citenamefont {Sauer},\ and\
  \citenamefont {Tarbutt}}]{Hinds2017}%
  \BibitemOpen
  \bibfield  {author} {\bibinfo {author} {\bibfnamefont {S.}~\bibnamefont
  {Truppe}}, \bibinfo {author} {\bibfnamefont {H.~J.}\ \bibnamefont
  {Williams}}, \bibinfo {author} {\bibfnamefont {M.}~\bibnamefont {Hambach}},
  \bibinfo {author} {\bibfnamefont {L.}~\bibnamefont {Caldwell}}, \bibinfo
  {author} {\bibfnamefont {N.~J.}\ \bibnamefont {Fitch}}, \bibinfo {author}
  {\bibfnamefont {E.~A.}\ \bibnamefont {Hinds}}, \bibinfo {author}
  {\bibfnamefont {B.~E.}\ \bibnamefont {Sauer}}, \ and\ \bibinfo {author}
  {\bibfnamefont {M.~R.}\ \bibnamefont {Tarbutt}},\ }\bibfield  {title}
  {\enquote {\bibinfo {title} {Molecules cooled below the {D}oppler limit},}\
  }\href@noop {} {\bibfield  {journal} {\bibinfo  {journal} {Nature Physics}\
  }\textbf {\bibinfo {volume} {13}},\ \bibinfo {pages} {1173} (\bibinfo {year}
  {2017})}\BibitemShut {NoStop}%
\bibitem [{\citenamefont {Collopy}\ \emph {et~al.}(2018)\citenamefont
  {Collopy}, \citenamefont {Ding}, \citenamefont {Wu}, \citenamefont
  {Finneran}, \citenamefont {Anderegg}, \citenamefont {Augenbraun},
  \citenamefont {Doyle},\ and\ \citenamefont {Ye}}]{Collopy2018}%
  \BibitemOpen
  \bibfield  {author} {\bibinfo {author} {\bibfnamefont {Alejandra~L.}\
  \bibnamefont {Collopy}}, \bibinfo {author} {\bibfnamefont {Shiqian}\
  \bibnamefont {Ding}}, \bibinfo {author} {\bibfnamefont {Yewei}\ \bibnamefont
  {Wu}}, \bibinfo {author} {\bibfnamefont {Ian~A.}\ \bibnamefont {Finneran}},
  \bibinfo {author} {\bibfnamefont {Lo\"{\i}c}\ \bibnamefont {Anderegg}},
  \bibinfo {author} {\bibfnamefont {Benjamin~L.}\ \bibnamefont {Augenbraun}},
  \bibinfo {author} {\bibfnamefont {John~M.}\ \bibnamefont {Doyle}}, \ and\
  \bibinfo {author} {\bibfnamefont {Jun}\ \bibnamefont {Ye}},\ }\bibfield
  {title} {\enquote {\bibinfo {title} {3d magneto-optical trap of yttrium
  monoxide},}\ }\href {\doibase 10.1103/PhysRevLett.121.213201} {\bibfield
  {journal} {\bibinfo  {journal} {Phys. Rev. Lett.}\ }\textbf {\bibinfo
  {volume} {121}},\ \bibinfo {pages} {213201} (\bibinfo {year}
  {2018})}\BibitemShut {NoStop}%
\bibitem [{\citenamefont {Lim}\ \emph {et~al.}(2018)\citenamefont {Lim},
  \citenamefont {Almond}, \citenamefont {Trigatzis}, \citenamefont {Devlin},
  \citenamefont {Fitch}, \citenamefont {Sauer}, \citenamefont {Tarbutt},\ and\
  \citenamefont {Hinds}}]{Lim2018}%
  \BibitemOpen
  \bibfield  {author} {\bibinfo {author} {\bibfnamefont {J.}~\bibnamefont
  {Lim}}, \bibinfo {author} {\bibfnamefont {J.~R.}\ \bibnamefont {Almond}},
  \bibinfo {author} {\bibfnamefont {M.~A.}\ \bibnamefont {Trigatzis}}, \bibinfo
  {author} {\bibfnamefont {J.~A.}\ \bibnamefont {Devlin}}, \bibinfo {author}
  {\bibfnamefont {N.~J.}\ \bibnamefont {Fitch}}, \bibinfo {author}
  {\bibfnamefont {B.~E.}\ \bibnamefont {Sauer}}, \bibinfo {author}
  {\bibfnamefont {M.~R.}\ \bibnamefont {Tarbutt}}, \ and\ \bibinfo {author}
  {\bibfnamefont {E.~A.}\ \bibnamefont {Hinds}},\ }\bibfield  {title} {\enquote
  {\bibinfo {title} {Laser cooled ybf molecules for measuring the electron's
  electric dipole moment},}\ }\href {\doibase 10.1103/PhysRevLett.120.123201}
  {\bibfield  {journal} {\bibinfo  {journal} {Phys. Rev. Lett.}\ }\textbf
  {\bibinfo {volume} {120}},\ \bibinfo {pages} {123201} (\bibinfo {year}
  {2018})}\BibitemShut {NoStop}%
\bibitem [{\citenamefont {Isaev}\ and\ \citenamefont
  {Berger}(2016)}]{Isaev2016}%
  \BibitemOpen
  \bibfield  {author} {\bibinfo {author} {\bibfnamefont {Timur~A.}\
  \bibnamefont {Isaev}}\ and\ \bibinfo {author} {\bibfnamefont {Robert}\
  \bibnamefont {Berger}},\ }\bibfield  {title} {\enquote {\bibinfo {title}
  {Polyatomic candidates for cooling of molecules with lasers from simple
  theoretical concepts},}\ }\href@noop {} {\bibfield  {journal} {\bibinfo
  {journal} {Phys. Rev. Lett.}\ }\textbf {\bibinfo {volume} {116}},\ \bibinfo
  {pages} {063006} (\bibinfo {year} {2016})}\BibitemShut {NoStop}%
\bibitem [{\citenamefont {Kozyryev}\ \emph {et~al.}(2017)\citenamefont
  {Kozyryev}, \citenamefont {Baum}, \citenamefont {Matsuda}, \citenamefont
  {Augenbraun}, \citenamefont {Anderegg}, \citenamefont {Sedlack},\ and\
  \citenamefont {Doyle}}]{Doyle2017}%
  \BibitemOpen
  \bibfield  {author} {\bibinfo {author} {\bibfnamefont {I.}~\bibnamefont
  {Kozyryev}}, \bibinfo {author} {\bibfnamefont {L.}~\bibnamefont {Baum}},
  \bibinfo {author} {\bibfnamefont {K.}~\bibnamefont {Matsuda}}, \bibinfo
  {author} {\bibfnamefont {B.~L.}\ \bibnamefont {Augenbraun}}, \bibinfo
  {author} {\bibfnamefont {L.}~\bibnamefont {Anderegg}}, \bibinfo {author}
  {\bibfnamefont {A.~P.}\ \bibnamefont {Sedlack}}, \ and\ \bibinfo {author}
  {\bibfnamefont {J.~M.}\ \bibnamefont {Doyle}},\ }\bibfield  {title} {\enquote
  {\bibinfo {title} {Sisyphus laser cooling of a polyatomic molecule},}\
  }\href@noop {} {\bibfield  {journal} {\bibinfo  {journal} {Phys. Rev. Lett.}\
  }\textbf {\bibinfo {volume} {118}},\ \bibinfo {pages} {173201} (\bibinfo
  {year} {2017})}\BibitemShut {NoStop}%
\bibitem [{\citenamefont {Kozyryev}\ \emph {et~al.}(2019)\citenamefont
  {Kozyryev}, \citenamefont {Steimle}, \citenamefont {Yu}, \citenamefont
  {Nguyen},\ and\ \citenamefont {Doyle}}]{Kosyryev_2019}%
  \BibitemOpen
  \bibfield  {author} {\bibinfo {author} {\bibfnamefont {Ivan}\ \bibnamefont
  {Kozyryev}}, \bibinfo {author} {\bibfnamefont {Timothy~C.}\ \bibnamefont
  {Steimle}}, \bibinfo {author} {\bibfnamefont {Phelan}\ \bibnamefont {Yu}},
  \bibinfo {author} {\bibfnamefont {Duc-Trung}\ \bibnamefont {Nguyen}}, \ and\
  \bibinfo {author} {\bibfnamefont {John~M.}\ \bibnamefont {Doyle}},\
  }\bibfield  {title} {\enquote {\bibinfo {title} {Determination of {CaOH} and
  {CaOCH}$_3$ vibrational branching ratios for direct laser cooling and
  trapping},}\ }\href@noop {} {\bibfield  {journal} {\bibinfo  {journal} {New
  J. Phys.}\ }\textbf {\bibinfo {volume} {21}},\ \bibinfo {pages} {052002}
  (\bibinfo {year} {2019})}\BibitemShut {NoStop}%
\bibitem [{\citenamefont {Klos}\ and\ \citenamefont
  {Kotochigova}(2019)}]{Klos2019}%
  \BibitemOpen
  \bibfield  {author} {\bibinfo {author} {\bibfnamefont {Jacek}\ \bibnamefont
  {Klos}}\ and\ \bibinfo {author} {\bibfnamefont {Svetlana}\ \bibnamefont
  {Kotochigova}},\ }\href@noop {} {\enquote {\bibinfo {title} {Prospects for
  laser cooling of polyatomic molecules with increasing complexity},}\ }
  (\bibinfo {year} {2019}),\ \bibinfo {note} {in preparation}\BibitemShut
  {NoStop}%
\bibitem [{\citenamefont {Kozyryev}\ \emph {et~al.}(2016)\citenamefont
  {Kozyryev}, \citenamefont {Baum}, \citenamefont {Matsuda},\ and\
  \citenamefont {Doyle}}]{Kozyryev2016}%
  \BibitemOpen
  \bibfield  {author} {\bibinfo {author} {\bibfnamefont {I.}~\bibnamefont
  {Kozyryev}}, \bibinfo {author} {\bibfnamefont {L.}~\bibnamefont {Baum}},
  \bibinfo {author} {\bibfnamefont {K.}~\bibnamefont {Matsuda}}, \ and\
  \bibinfo {author} {\bibfnamefont {J.~M.}\ \bibnamefont {Doyle}},\ }\bibfield
  {title} {\enquote {\bibinfo {title} {Proposal for laser cooling of complex
  polyatomic molecules},}\ }\href@noop {} {\bibfield  {journal} {\bibinfo
  {journal} {Chem. Phys. Chem.}\ }\textbf {\bibinfo {volume} {17}},\ \bibinfo
  {pages} {3641--3648} (\bibinfo {year} {2016})}\BibitemShut {NoStop}%
\bibitem [{\citenamefont {Yarkony}(1996)}]{RennerTeller}%
  \BibitemOpen
  \bibfield  {author} {\bibinfo {author} {\bibfnamefont {D.~R.}\ \bibnamefont
  {Yarkony}},\ }\bibfield  {title} {\enquote {\bibinfo {title} {Diabolical
  conical intersections},}\ }\href@noop {} {\bibfield  {journal} {\bibinfo
  {journal} {Rev. Mod. Phys.}\ }\textbf {\bibinfo {volume} {68}},\ \bibinfo
  {pages} {985--1013} (\bibinfo {year} {1996})}\BibitemShut {NoStop}%
\bibitem [{\citenamefont {Kas}\ \emph {et~al.}(2017)\citenamefont {Kas},
  \citenamefont {Loreau}, \citenamefont {Lievin},\ and\ \citenamefont
  {Vaeck}}]{Vaeck2017}%
  \BibitemOpen
  \bibfield  {author} {\bibinfo {author} {\bibfnamefont {M.}~\bibnamefont
  {Kas}}, \bibinfo {author} {\bibfnamefont {J.}~\bibnamefont {Loreau}},
  \bibinfo {author} {\bibfnamefont {J.}~\bibnamefont {Lievin}}, \ and\ \bibinfo
  {author} {\bibfnamefont {N.}~\bibnamefont {Vaeck}},\ }\bibfield  {title}
  {\enquote {\bibinfo {title} {{\em Ab initio} study of the neutral and anionic
  alkali and alkaline earth hydroxides: {E}lectronic structure and prospects
  for sympathetic cooling of {OH}$^-$},}\ }\href@noop {} {\bibfield  {journal}
  {\bibinfo  {journal} {J. Chem. Phys.}\ }\textbf {\bibinfo {volume} {146}},\
  \bibinfo {pages} {194309} (\bibinfo {year} {2017})}\BibitemShut {NoStop}%
\bibitem [{\citenamefont {Theodorakopoulos}\ \emph {et~al.}(1999)\citenamefont
  {Theodorakopoulos}, \citenamefont {Petsalakis},\ and\ \citenamefont
  {Hamilton}}]{Pesalakis1999}%
  \BibitemOpen
  \bibfield  {author} {\bibinfo {author} {\bibfnamefont {Giannoula}\
  \bibnamefont {Theodorakopoulos}}, \bibinfo {author} {\bibfnamefont {Ioannis}\
  \bibnamefont {Petsalakis}}, \ and\ \bibinfo {author} {\bibfnamefont {Ian}\
  \bibnamefont {Hamilton}},\ }\bibfield  {title} {\enquote {\bibinfo {title}
  {{\em Ab initio} calculations on the ground and excited states of {BeOH} and
  {MgOH}},}\ }\href@noop {} {\bibfield  {journal} {\bibinfo  {journal} {J.
  Chem. Phys.}\ }\textbf {\bibinfo {volume} {111}},\ \bibinfo {pages}
  {10484--10490} (\bibinfo {year} {1999})}\BibitemShut {NoStop}%
\bibitem [{\citenamefont {Presunka}\ and\ \citenamefont
  {Coxon}(1994)}]{Presunka:1994}%
  \BibitemOpen
  \bibfield  {author} {\bibinfo {author} {\bibfnamefont {Paul~I.}\ \bibnamefont
  {Presunka}}\ and\ \bibinfo {author} {\bibfnamefont {John~A.}\ \bibnamefont
  {Coxon}},\ }\bibfield  {title} {\enquote {\bibinfo {title} {Laser
  spectroscopy of the \~{A}$^2{\Pi}$ - \~{X}$^2{\Sigma}^+$ transition of
  {SrOH}: Deperturbation analysis of {K}‐resonance in the v$_2$=1 level of
  the \~{A}$^2{\Pi}$},}\ }\href@noop {} {\bibfield  {journal} {\bibinfo
  {journal} {J. Chem. Phys.}\ }\textbf {\bibinfo {volume} {101}},\ \bibinfo
  {pages} {201--222} (\bibinfo {year} {1994})}\BibitemShut {NoStop}%
\bibitem [{\citenamefont {Nguyen}\ \emph {et~al.}(2018)\citenamefont {Nguyen},
  \citenamefont {Steimle}, \citenamefont {Kozyryev}, \citenamefont {Huang},\
  and\ \citenamefont {McCoy}}]{Nguyen:2018}%
  \BibitemOpen
  \bibfield  {author} {\bibinfo {author} {\bibfnamefont {Duc-Trung}\
  \bibnamefont {Nguyen}}, \bibinfo {author} {\bibfnamefont {Timothy~C.}\
  \bibnamefont {Steimle}}, \bibinfo {author} {\bibfnamefont {Ivan}\
  \bibnamefont {Kozyryev}}, \bibinfo {author} {\bibfnamefont {Meng}\
  \bibnamefont {Huang}}, \ and\ \bibinfo {author} {\bibfnamefont {Anne~B.}\
  \bibnamefont {McCoy}},\ }\bibfield  {title} {\enquote {\bibinfo {title}
  {Fluorescence branching ratios and magnetic tuning of the visible spectrum of
  {SrOH}},}\ }\href@noop {} {\bibfield  {journal} {\bibinfo  {journal} {J. Mol.
  Spec.}\ }\textbf {\bibinfo {volume} {347}},\ \bibinfo {pages} {7 -- 18}
  (\bibinfo {year} {2018})}\BibitemShut {NoStop}%
\bibitem [{\citenamefont {Berry}(1984)}]{Berry:1984}%
  \BibitemOpen
  \bibfield  {author} {\bibinfo {author} {\bibfnamefont {M.~V.}\ \bibnamefont
  {Berry}},\ }\bibfield  {title} {\enquote {\bibinfo {title} {Quantum phase
  factors accompanying adiabatic changes},}\ }\href@noop {} {\bibfield
  {journal} {\bibinfo  {journal} {Proc. R. Soc. Lond. A}\ }\textbf {\bibinfo
  {volume} {392}},\ \bibinfo {pages} {45--57} (\bibinfo {year}
  {1984})}\BibitemShut {NoStop}%
\bibitem [{\citenamefont {Xie}\ \emph {et~al.}(2017)\citenamefont {Xie},
  \citenamefont {Kendrick}, \citenamefont {Yarkony},\ and\ \citenamefont
  {Guo}}]{Guo:2017}%
  \BibitemOpen
  \bibfield  {author} {\bibinfo {author} {\bibfnamefont {Changjian}\
  \bibnamefont {Xie}}, \bibinfo {author} {\bibfnamefont {Brian~K.}\
  \bibnamefont {Kendrick}}, \bibinfo {author} {\bibfnamefont {David~R.}\
  \bibnamefont {Yarkony}}, \ and\ \bibinfo {author} {\bibfnamefont {Huo}\
  \bibnamefont {Guo}},\ }\bibfield  {title} {\enquote {\bibinfo {title}
  {Constructive and destructive interference in nonadiabatic tunneling via
  conical intersection},}\ }\href@noop {} {\bibfield  {journal} {\bibinfo
  {journal} {J. Chem. Theory Comput}\ }\textbf {\bibinfo {volume} {13}},\
  \bibinfo {pages} {1902--1910} (\bibinfo {year} {2017})}\BibitemShut {NoStop}%
\bibitem [{\citenamefont {Born}\ and\ \citenamefont {Oppenheimer}(1927)}]{BO}%
  \BibitemOpen
  \bibfield  {author} {\bibinfo {author} {\bibfnamefont {M.}~\bibnamefont
  {Born}}\ and\ \bibinfo {author} {\bibfnamefont {R.}~\bibnamefont
  {Oppenheimer}},\ }\bibfield  {title} {\enquote {\bibinfo {title} {Zur
  quantentheorie der molekeln},}\ }\href@noop {} {\bibfield  {journal}
  {\bibinfo  {journal} {Ann. Phys.}\ }\textbf {\bibinfo {volume} {84}},\
  \bibinfo {pages} {457} (\bibinfo {year} {1927})}\BibitemShut {NoStop}%
\bibitem [{\citenamefont {Domcke}\ \emph {et~al.}(2004)\citenamefont {Domcke},
  \citenamefont {Yarkony},\ and\ \citenamefont {K{\"o}ppel}}]{Domcke2004}%
  \BibitemOpen
  \bibfield  {author} {\bibinfo {author} {\bibfnamefont {W.}~\bibnamefont
  {Domcke}}, \bibinfo {author} {\bibfnamefont {D.}~\bibnamefont {Yarkony}}, \
  and\ \bibinfo {author} {\bibfnamefont {H.}~\bibnamefont {K{\"o}ppel}},\
  }\href@noop {} {\emph {\bibinfo {title} {Conical Intersections: Electronic
  Structure, Dynamics \& Spectroscopy}}},\ Advanced series in physical
  chemistry\ (\bibinfo  {publisher} {World Scientific, Singapore},\ \bibinfo
  {year} {2004})\BibitemShut {NoStop}%
\bibitem [{\citenamefont {Nikitin}(2006)}]{Nikitin2006Drake}%
  \BibitemOpen
  \bibfield  {author} {\bibinfo {author} {\bibfnamefont {Evgueni~E.}\
  \bibnamefont {Nikitin}},\ }\bibfield  {title} {\enquote {\bibinfo {title}
  {Adiabatic and diabatic collision processes at low energies},}\ }in\
  \href@noop {} {\emph {\bibinfo {booktitle} {Springer Handbooks of Atomic,
  Molecular, and Optical Physics}}},\ \bibinfo {editor} {edited by\ \bibinfo
  {editor} {\bibfnamefont {Gordon W.~F.}\ \bibnamefont {Drake}}}\ (\bibinfo
  {publisher} {Springer},\ \bibinfo {address} {New York},\ \bibinfo {year}
  {2006})\ Chap.~\bibinfo {chapter} {49}, pp.\ \bibinfo {pages}
  {741--752}\BibitemShut {NoStop}%
\bibitem [{\citenamefont {Akima}(1970)}]{AkimaIn}%
  \BibitemOpen
  \bibfield  {author} {\bibinfo {author} {\bibfnamefont {Hiroshi}\ \bibnamefont
  {Akima}},\ }\bibfield  {title} {\enquote {\bibinfo {title} {A new method of
  interpolation and smooth curve fitting based on local procedures},}\
  }\href@noop {} {\bibfield  {journal} {\bibinfo  {journal} {J. ACM}\ }\textbf
  {\bibinfo {volume} {17}},\ \bibinfo {pages} {589--602} (\bibinfo {year}
  {1970})}\BibitemShut {NoStop}%
\bibitem [{\citenamefont {Renner}(1934)}]{Renner1934}%
  \BibitemOpen
  \bibfield  {author} {\bibinfo {author} {\bibfnamefont {R.}~\bibnamefont
  {Renner}},\ }\bibfield  {title} {\enquote {\bibinfo {title} {Zur theorie der
  wechselwirkung zwischen elektronen- und kernbewegung bei dreiatomigen,
  stabf{\"o}rmigen molek{\"u}len},}\ }\href@noop {} {\bibfield  {journal}
  {\bibinfo  {journal} {Zeitschrift f{\"u}r Physik}\ }\textbf {\bibinfo
  {volume} {92}},\ \bibinfo {pages} {172--193} (\bibinfo {year}
  {1934})}\BibitemShut {NoStop}%
\bibitem [{\citenamefont {Herzberg}\ and\ \citenamefont
  {Teller}(1933)}]{Teller1933}%
  \BibitemOpen
  \bibfield  {author} {\bibinfo {author} {\bibfnamefont {G.}~\bibnamefont
  {Herzberg}}\ and\ \bibinfo {author} {\bibfnamefont {E.}~\bibnamefont
  {Teller}},\ }\bibfield  {title} {\enquote {\bibinfo {title}
  {Schwingungsstruktur der elektronen{\"u}berg{\"a}nge bei mehratomigen
  molek{\"u}len},}\ }\href@noop {} {\bibfield  {journal} {\bibinfo  {journal}
  {Zeitschrift f{\"u}r Physikalische Chemie}\ }\textbf {\bibinfo {volume}
  {21B}},\ \bibinfo {pages} {410--446} (\bibinfo {year} {1933})}\BibitemShut
  {NoStop}%
\bibitem [{\citenamefont {Stanton}\ \emph {et~al.}()\citenamefont {Stanton},
  \citenamefont {Gauss}, \citenamefont {Cheng}, \citenamefont {Harding},
  \citenamefont {Matthews},\ and\ \citenamefont {Szalay}}]{cfour}%
  \BibitemOpen
  \bibfield  {author} {\bibinfo {author} {\bibfnamefont {J.~F.}\ \bibnamefont
  {Stanton}}, \bibinfo {author} {\bibfnamefont {J.}~\bibnamefont {Gauss}},
  \bibinfo {author} {\bibfnamefont {L.}~\bibnamefont {Cheng}}, \bibinfo
  {author} {\bibfnamefont {M.~E.}\ \bibnamefont {Harding}}, \bibinfo {author}
  {\bibfnamefont {D.~A.}\ \bibnamefont {Matthews}}, \ and\ \bibinfo {author}
  {\bibfnamefont {P.~G.}\ \bibnamefont {Szalay}},\ }\href@noop {} {\enquote
  {\bibinfo {title} {{{CFOUR}, Coupled-Cluster techniques for Computational
  Chemistry, a quantum-chemical program package}},}\ }\bibinfo {note} {{W}ith
  contributions from {A}.{A}. {A}uer, {R}.{J}. {B}artlett, {U}. {B}enedikt,
  {C}. {B}erger, {D}.{E}. {B}ernholdt, {Y}.{J}. {B}omble, {O}. {C}hristiansen,
  {F}. Engel, {R}. Faber, {M}. {H}eckert, {O}. {H}eun, {M}. Hilgenberg, {C}.
  {H}uber, {T}.-{C}. {J}agau, {D}. {J}onsson, {J}. {J}us{\'e}lius, {T}. Kirsch,
  {K}. {K}lein, {W}.{J}. {L}auderdale, {F}. {L}ipparini, {T}. {M}etzroth,
  {L}.{A}. {M}{\"u}ck, {D}.{P}. {O}'{N}eill, {D}.{R}. {P}rice, {E}. {P}rochnow,
  {C}. {P}uzzarini, {K}. {R}uud, {F}. {S}chiffmann, {W}. {S}chwalbach, {C}.
  {S}immons, {S}. {S}topkowicz, {A}. {T}ajti, {J}. {V}{\'a}zquez, {F}. {W}ang,
  {J}.{D}. {W}atts and the integral packages {MOLECULE} ({J}. {A}lml{\"o}f and
  {P}.{R}. {T}aylor), {PROPS} ({P}.{R}. {T}aylor), {ABACUS} ({T}. {H}elgaker,
  {H}.{J}. {A}a. {J}ensen, {P}. {J}{\o}rgensen, and {J}. {O}lsen), and {ECP}
  routines by {A}. {V}. {M}itin and {C}. van {W}{\"u}llen. {F}or the current
  version, see http://www.cfour.de.}\BibitemShut {Stop}%
\bibitem [{\citenamefont {Weigend}\ and\ \citenamefont
  {Ahlrichs}(2005)}]{def2-QZVPP}%
  \BibitemOpen
  \bibfield  {author} {\bibinfo {author} {\bibfnamefont {F.}~\bibnamefont
  {Weigend}}\ and\ \bibinfo {author} {\bibfnamefont {R.}~\bibnamefont
  {Ahlrichs}},\ }\bibfield  {title} {\enquote {\bibinfo {title} {Balanced basis
  sets of split valence{,} triple zeta valence and quadruple zeta valence
  quality for {H} to {R}n: {D}esign and assessment of accuracy},}\ }\href@noop
  {} {\bibfield  {journal} {\bibinfo  {journal} {Phys. Chem. Chem. Phys.}\
  }\textbf {\bibinfo {volume} {7}},\ \bibinfo {pages} {3297--3305} (\bibinfo
  {year} {2005})}\BibitemShut {NoStop}%
\bibitem [{\citenamefont {Lim}\ \emph {et~al.}(2006)\citenamefont {Lim},
  \citenamefont {Stoll},\ and\ \citenamefont {Schwerdtfeger}}]{Lim:2006}%
  \BibitemOpen
  \bibfield  {author} {\bibinfo {author} {\bibfnamefont {Ivan~S.}\ \bibnamefont
  {Lim}}, \bibinfo {author} {\bibfnamefont {Hermann}\ \bibnamefont {Stoll}}, \
  and\ \bibinfo {author} {\bibfnamefont {Peter}\ \bibnamefont
  {Schwerdtfeger}},\ }\bibfield  {title} {\enquote {\bibinfo {title}
  {Relativistic small-core energy-consistent pseudopotentials for the
  alkaline-earth elements from {C}a to {R}a},}\ }\href@noop {} {\bibfield
  {journal} {\bibinfo  {journal} {J. Chem. Phys.}\ }\textbf {\bibinfo {volume}
  {124}},\ \bibinfo {pages} {034107} (\bibinfo {year} {2006})}\BibitemShut
  {NoStop}%
\bibitem [{\citenamefont {Dunning}(1989)}]{dunning:89}%
  \BibitemOpen
  \bibfield  {author} {\bibinfo {author} {\bibfnamefont {Thom~H.}\ \bibnamefont
  {Dunning}},\ }\bibfield  {title} {\enquote {\bibinfo {title} {Gaussian basis
  sets for use in correlated molecular calculations. i. {T}he atoms boron
  through neon and hydrogen},}\ }\href@noop {} {\bibfield  {journal} {\bibinfo
  {journal} {J. Chem. Phys.}\ }\textbf {\bibinfo {volume} {90}},\ \bibinfo
  {pages} {1007--1023} (\bibinfo {year} {1989})}\BibitemShut {NoStop}%
\bibitem [{\citenamefont {Shao}\ \emph {et~al.}(2015)\citenamefont {Shao},
  \citenamefont {Gan}, \citenamefont {Epifanovsky}, \citenamefont {Gilbert},
  \citenamefont {Wormit}, \citenamefont {Kussmann}, \citenamefont {Lange},
  \citenamefont {Behn}, \citenamefont {Deng}, \citenamefont {Feng},
  \citenamefont {Ghosh}, \citenamefont {Goldey}, \citenamefont {Horn},
  \citenamefont {Jacobson}, \citenamefont {Kaliman}, \citenamefont
  {Khaliullin}, \citenamefont {Ku{\'s}}, \citenamefont {Landau}, \citenamefont
  {Liu}, \citenamefont {Proynov}, \citenamefont {Rhee}, \citenamefont
  {Richard}, \citenamefont {Rohrdanz}, \citenamefont {Steele}, \citenamefont
  {Sundstrom}, \citenamefont {III}, \citenamefont {Zimmerman}, \citenamefont
  {Zuev}, \citenamefont {Albrecht}, \citenamefont {Alguire}, \citenamefont
  {Austin}, \citenamefont {Beran}, \citenamefont {Bernard}, \citenamefont
  {Berquist}, \citenamefont {Brandhorst}, \citenamefont {Bravaya},
  \citenamefont {Brown}, \citenamefont {Casanova}, \citenamefont {Chang},
  \citenamefont {Chen}, \citenamefont {Chien}, \citenamefont {Closser},
  \citenamefont {Crittenden}, \citenamefont {Diedenhofen}, \citenamefont {Jr.},
  \citenamefont {Do}, \citenamefont {Dutoi}, \citenamefont {Edgar},
  \citenamefont {Fatehi}, \citenamefont {Fusti-Molnar}, \citenamefont
  {Ghysels}, \citenamefont {Golubeva-Zadorozhnaya}, \citenamefont {Gomes},
  \citenamefont {Hanson-Heine}, \citenamefont {Harbach}, \citenamefont
  {Hauser}, \citenamefont {Hohenstein}, \citenamefont {Holden}, \citenamefont
  {Jagau}, \citenamefont {Ji}, \citenamefont {Kaduk}, \citenamefont
  {Khistyaev}, \citenamefont {Kim}, \citenamefont {Kim}, \citenamefont {King},
  \citenamefont {Klunzinger}, \citenamefont {Kosenkov}, \citenamefont
  {Kowalczyk}, \citenamefont {Krauter}, \citenamefont {Lao}, \citenamefont
  {Laurent}, \citenamefont {Lawler}, \citenamefont {Levchenko}, \citenamefont
  {Lin}, \citenamefont {Liu}, \citenamefont {Livshits}, \citenamefont {Lochan},
  \citenamefont {Luenser}, \citenamefont {Manohar}, \citenamefont {Manzer},
  \citenamefont {Mao}, \citenamefont {Mardirossian}, \citenamefont {Marenich},
  \citenamefont {Maurer}, \citenamefont {Mayhall}, \citenamefont {Neuscamman},
  \citenamefont {Oana}, \citenamefont {Olivares-Amaya}, \citenamefont
  {O'Neill}, \citenamefont {Parkhill}, \citenamefont {Perrine}, \citenamefont
  {Peverati}, \citenamefont {Prociuk}, \citenamefont {Rehn}, \citenamefont
  {Rosta}, \citenamefont {Russ}, \citenamefont {Sharada}, \citenamefont
  {Sharma}, \citenamefont {Small}, \citenamefont {Sodt}, \citenamefont {Stein},
  \citenamefont {St{\"u}ck}, \citenamefont {Su}, \citenamefont {Thom},
  \citenamefont {Tsuchimochi}, \citenamefont {Vanovschi}, \citenamefont {Vogt},
  \citenamefont {Vydrov}, \citenamefont {Wang}, \citenamefont {Watson},
  \citenamefont {Wenzel}, \citenamefont {White}, \citenamefont {Williams},
  \citenamefont {Yang}, \citenamefont {Yeganeh}, \citenamefont {Yost},
  \citenamefont {You}, \citenamefont {Zhang}, \citenamefont {Zhang},
  \citenamefont {Zhao}, \citenamefont {Brooks}, \citenamefont {Chan},
  \citenamefont {Chipman}, \citenamefont {Cramer}, \citenamefont {III},
  \citenamefont {Gordon}, \citenamefont {Hehre}, \citenamefont {Klamt},
  \citenamefont {III}, \citenamefont {Schmidt}, \citenamefont {Sherrill},
  \citenamefont {Truhlar}, \citenamefont {Warshel}, \citenamefont {Xu},
  \citenamefont {Aspuru-Guzik}, \citenamefont {Baer}, \citenamefont {Bell},
  \citenamefont {Besley}, \citenamefont {Chai}, \citenamefont {Dreuw},
  \citenamefont {Dunietz}, \citenamefont {Furlani}, \citenamefont {Gwaltney},
  \citenamefont {Hsu}, \citenamefont {Jung}, \citenamefont {Kong},
  \citenamefont {Lambrecht}, \citenamefont {Liang}, \citenamefont {Ochsenfeld},
  \citenamefont {Rassolov}, \citenamefont {Slipchenko}, \citenamefont
  {Subotnik}, \citenamefont {Voorhis}, \citenamefont {Herbert}, \citenamefont
  {Krylov}, \citenamefont {Gill},\ and\ \citenamefont {Head-Gordon}}]{QChem}%
  \BibitemOpen
  \bibfield  {author} {\bibinfo {author} {\bibfnamefont {Yihan}\ \bibnamefont
  {Shao}}, \bibinfo {author} {\bibfnamefont {Zhengting}\ \bibnamefont {Gan}},
  \bibinfo {author} {\bibfnamefont {Evgeny}\ \bibnamefont {Epifanovsky}},
  \bibinfo {author} {\bibfnamefont {Andrew~T.B.}\ \bibnamefont {Gilbert}},
  \bibinfo {author} {\bibfnamefont {Michael}\ \bibnamefont {Wormit}}, \bibinfo
  {author} {\bibfnamefont {Joerg}\ \bibnamefont {Kussmann}}, \bibinfo {author}
  {\bibfnamefont {Adrian~W.}\ \bibnamefont {Lange}}, \bibinfo {author}
  {\bibfnamefont {Andrew}\ \bibnamefont {Behn}}, \bibinfo {author}
  {\bibfnamefont {Jia}\ \bibnamefont {Deng}}, \bibinfo {author} {\bibfnamefont
  {Xintian}\ \bibnamefont {Feng}}, \bibinfo {author} {\bibfnamefont
  {Debashree}\ \bibnamefont {Ghosh}}, \bibinfo {author} {\bibfnamefont
  {Matthew}\ \bibnamefont {Goldey}}, \bibinfo {author} {\bibfnamefont
  {Paul~R.}\ \bibnamefont {Horn}}, \bibinfo {author} {\bibfnamefont {Leif~D.}\
  \bibnamefont {Jacobson}}, \bibinfo {author} {\bibfnamefont {Ilya}\
  \bibnamefont {Kaliman}}, \bibinfo {author} {\bibfnamefont {Rustam~Z.}\
  \bibnamefont {Khaliullin}}, \bibinfo {author} {\bibfnamefont {Tomasz}\
  \bibnamefont {Ku{\'s}}}, \bibinfo {author} {\bibfnamefont {Arie}\
  \bibnamefont {Landau}}, \bibinfo {author} {\bibfnamefont {Jie}\ \bibnamefont
  {Liu}}, \bibinfo {author} {\bibfnamefont {Emil~I.}\ \bibnamefont {Proynov}},
  \bibinfo {author} {\bibfnamefont {Young~Min}\ \bibnamefont {Rhee}}, \bibinfo
  {author} {\bibfnamefont {Ryan~M.}\ \bibnamefont {Richard}}, \bibinfo {author}
  {\bibfnamefont {Mary~A.}\ \bibnamefont {Rohrdanz}}, \bibinfo {author}
  {\bibfnamefont {Ryan~P.}\ \bibnamefont {Steele}}, \bibinfo {author}
  {\bibfnamefont {Eric~J.}\ \bibnamefont {Sundstrom}}, \bibinfo {author}
  {\bibfnamefont {H.~Lee~Woodcock}\ \bibnamefont {III}}, \bibinfo {author}
  {\bibfnamefont {Paul~M.}\ \bibnamefont {Zimmerman}}, \bibinfo {author}
  {\bibfnamefont {Dmitry}\ \bibnamefont {Zuev}}, \bibinfo {author}
  {\bibfnamefont {Ben}\ \bibnamefont {Albrecht}}, \bibinfo {author}
  {\bibfnamefont {Ethan}\ \bibnamefont {Alguire}}, \bibinfo {author}
  {\bibfnamefont {Brian}\ \bibnamefont {Austin}}, \bibinfo {author}
  {\bibfnamefont {Gregory J.~O.}\ \bibnamefont {Beran}}, \bibinfo {author}
  {\bibfnamefont {Yves~A.}\ \bibnamefont {Bernard}}, \bibinfo {author}
  {\bibfnamefont {Eric}\ \bibnamefont {Berquist}}, \bibinfo {author}
  {\bibfnamefont {Kai}\ \bibnamefont {Brandhorst}}, \bibinfo {author}
  {\bibfnamefont {Ksenia~B.}\ \bibnamefont {Bravaya}}, \bibinfo {author}
  {\bibfnamefont {Shawn~T.}\ \bibnamefont {Brown}}, \bibinfo {author}
  {\bibfnamefont {David}\ \bibnamefont {Casanova}}, \bibinfo {author}
  {\bibfnamefont {Chun-Min}\ \bibnamefont {Chang}}, \bibinfo {author}
  {\bibfnamefont {Yunqing}\ \bibnamefont {Chen}}, \bibinfo {author}
  {\bibfnamefont {Siu~Hung}\ \bibnamefont {Chien}}, \bibinfo {author}
  {\bibfnamefont {Kristina~D.}\ \bibnamefont {Closser}}, \bibinfo {author}
  {\bibfnamefont {Deborah~L.}\ \bibnamefont {Crittenden}}, \bibinfo {author}
  {\bibfnamefont {Michael}\ \bibnamefont {Diedenhofen}}, \bibinfo {author}
  {\bibfnamefont {Robert A.~DiStasio}\ \bibnamefont {Jr.}}, \bibinfo {author}
  {\bibfnamefont {Hainam}\ \bibnamefont {Do}}, \bibinfo {author} {\bibfnamefont
  {Anthony~D.}\ \bibnamefont {Dutoi}}, \bibinfo {author} {\bibfnamefont
  {Richard~G.}\ \bibnamefont {Edgar}}, \bibinfo {author} {\bibfnamefont
  {Shervin}\ \bibnamefont {Fatehi}}, \bibinfo {author} {\bibfnamefont {Laszlo}\
  \bibnamefont {Fusti-Molnar}}, \bibinfo {author} {\bibfnamefont
  {An}~\bibnamefont {Ghysels}}, \bibinfo {author} {\bibfnamefont {Anna}\
  \bibnamefont {Golubeva-Zadorozhnaya}}, \bibinfo {author} {\bibfnamefont
  {Joseph}\ \bibnamefont {Gomes}}, \bibinfo {author} {\bibfnamefont
  {Magnus~W.D.}\ \bibnamefont {Hanson-Heine}}, \bibinfo {author} {\bibfnamefont
  {Philipp~H.P.}\ \bibnamefont {Harbach}}, \bibinfo {author} {\bibfnamefont
  {Andreas~W.}\ \bibnamefont {Hauser}}, \bibinfo {author} {\bibfnamefont
  {Edward~G.}\ \bibnamefont {Hohenstein}}, \bibinfo {author} {\bibfnamefont
  {Zachary~C.}\ \bibnamefont {Holden}}, \bibinfo {author} {\bibfnamefont
  {Thomas-C.}\ \bibnamefont {Jagau}}, \bibinfo {author} {\bibfnamefont
  {Hyunjun}\ \bibnamefont {Ji}}, \bibinfo {author} {\bibfnamefont {Benjamin}\
  \bibnamefont {Kaduk}}, \bibinfo {author} {\bibfnamefont {Kirill}\
  \bibnamefont {Khistyaev}}, \bibinfo {author} {\bibfnamefont {Jaehoon}\
  \bibnamefont {Kim}}, \bibinfo {author} {\bibfnamefont {Jihan}\ \bibnamefont
  {Kim}}, \bibinfo {author} {\bibfnamefont {Rollin~A.}\ \bibnamefont {King}},
  \bibinfo {author} {\bibfnamefont {Phil}\ \bibnamefont {Klunzinger}}, \bibinfo
  {author} {\bibfnamefont {Dmytro}\ \bibnamefont {Kosenkov}}, \bibinfo {author}
  {\bibfnamefont {Tim}\ \bibnamefont {Kowalczyk}}, \bibinfo {author}
  {\bibfnamefont {Caroline~M.}\ \bibnamefont {Krauter}}, \bibinfo {author}
  {\bibfnamefont {Ka~Un}\ \bibnamefont {Lao}}, \bibinfo {author} {\bibfnamefont
  {Ad{\`e}le~D.}\ \bibnamefont {Laurent}}, \bibinfo {author} {\bibfnamefont
  {Keith~V.}\ \bibnamefont {Lawler}}, \bibinfo {author} {\bibfnamefont
  {Sergey~V.}\ \bibnamefont {Levchenko}}, \bibinfo {author} {\bibfnamefont
  {Ching~Yeh}\ \bibnamefont {Lin}}, \bibinfo {author} {\bibfnamefont {Fenglai}\
  \bibnamefont {Liu}}, \bibinfo {author} {\bibfnamefont {Ester}\ \bibnamefont
  {Livshits}}, \bibinfo {author} {\bibfnamefont {Rohini~C.}\ \bibnamefont
  {Lochan}}, \bibinfo {author} {\bibfnamefont {Arne}\ \bibnamefont {Luenser}},
  \bibinfo {author} {\bibfnamefont {Prashant}\ \bibnamefont {Manohar}},
  \bibinfo {author} {\bibfnamefont {Samuel~F.}\ \bibnamefont {Manzer}},
  \bibinfo {author} {\bibfnamefont {Shan-Ping}\ \bibnamefont {Mao}}, \bibinfo
  {author} {\bibfnamefont {Narbe}\ \bibnamefont {Mardirossian}}, \bibinfo
  {author} {\bibfnamefont {Aleksandr~V.}\ \bibnamefont {Marenich}}, \bibinfo
  {author} {\bibfnamefont {Simon~A.}\ \bibnamefont {Maurer}}, \bibinfo {author}
  {\bibfnamefont {Nicholas~J.}\ \bibnamefont {Mayhall}}, \bibinfo {author}
  {\bibfnamefont {Eric}\ \bibnamefont {Neuscamman}}, \bibinfo {author}
  {\bibfnamefont {C.~Melania}\ \bibnamefont {Oana}}, \bibinfo {author}
  {\bibfnamefont {Roberto}\ \bibnamefont {Olivares-Amaya}}, \bibinfo {author}
  {\bibfnamefont {Darragh~P.}\ \bibnamefont {O'Neill}}, \bibinfo {author}
  {\bibfnamefont {John~A.}\ \bibnamefont {Parkhill}}, \bibinfo {author}
  {\bibfnamefont {Trilisa~M.}\ \bibnamefont {Perrine}}, \bibinfo {author}
  {\bibfnamefont {Roberto}\ \bibnamefont {Peverati}}, \bibinfo {author}
  {\bibfnamefont {Alexander}\ \bibnamefont {Prociuk}}, \bibinfo {author}
  {\bibfnamefont {Dirk~R.}\ \bibnamefont {Rehn}}, \bibinfo {author}
  {\bibfnamefont {Edina}\ \bibnamefont {Rosta}}, \bibinfo {author}
  {\bibfnamefont {Nicholas~J.}\ \bibnamefont {Russ}}, \bibinfo {author}
  {\bibfnamefont {Shaama~M.}\ \bibnamefont {Sharada}}, \bibinfo {author}
  {\bibfnamefont {Sandeep}\ \bibnamefont {Sharma}}, \bibinfo {author}
  {\bibfnamefont {David~W.}\ \bibnamefont {Small}}, \bibinfo {author}
  {\bibfnamefont {Alexander}\ \bibnamefont {Sodt}}, \bibinfo {author}
  {\bibfnamefont {Tamar}\ \bibnamefont {Stein}}, \bibinfo {author}
  {\bibfnamefont {David}\ \bibnamefont {St{\"u}ck}}, \bibinfo {author}
  {\bibfnamefont {Yu-Chuan}\ \bibnamefont {Su}}, \bibinfo {author}
  {\bibfnamefont {Alex~J.W.}\ \bibnamefont {Thom}}, \bibinfo {author}
  {\bibfnamefont {Takashi}\ \bibnamefont {Tsuchimochi}}, \bibinfo {author}
  {\bibfnamefont {Vitalii}\ \bibnamefont {Vanovschi}}, \bibinfo {author}
  {\bibfnamefont {Leslie}\ \bibnamefont {Vogt}}, \bibinfo {author}
  {\bibfnamefont {Oleg}\ \bibnamefont {Vydrov}}, \bibinfo {author}
  {\bibfnamefont {Tao}\ \bibnamefont {Wang}}, \bibinfo {author} {\bibfnamefont
  {Mark~A.}\ \bibnamefont {Watson}}, \bibinfo {author} {\bibfnamefont {Jan}\
  \bibnamefont {Wenzel}}, \bibinfo {author} {\bibfnamefont {Alec}\ \bibnamefont
  {White}}, \bibinfo {author} {\bibfnamefont {Christopher~F.}\ \bibnamefont
  {Williams}}, \bibinfo {author} {\bibfnamefont {Jun}\ \bibnamefont {Yang}},
  \bibinfo {author} {\bibfnamefont {Sina}\ \bibnamefont {Yeganeh}}, \bibinfo
  {author} {\bibfnamefont {Shane~R.}\ \bibnamefont {Yost}}, \bibinfo {author}
  {\bibfnamefont {Zhi-Qiang}\ \bibnamefont {You}}, \bibinfo {author}
  {\bibfnamefont {Igor~Ying}\ \bibnamefont {Zhang}}, \bibinfo {author}
  {\bibfnamefont {Xing}\ \bibnamefont {Zhang}}, \bibinfo {author}
  {\bibfnamefont {Yan}\ \bibnamefont {Zhao}}, \bibinfo {author} {\bibfnamefont
  {Bernard~R.}\ \bibnamefont {Brooks}}, \bibinfo {author} {\bibfnamefont
  {Garnet~K.L.}\ \bibnamefont {Chan}}, \bibinfo {author} {\bibfnamefont
  {Daniel~M.}\ \bibnamefont {Chipman}}, \bibinfo {author} {\bibfnamefont
  {Christopher~J.}\ \bibnamefont {Cramer}}, \bibinfo {author} {\bibfnamefont
  {William A.~Goddard}\ \bibnamefont {III}}, \bibinfo {author} {\bibfnamefont
  {Mark~S.}\ \bibnamefont {Gordon}}, \bibinfo {author} {\bibfnamefont
  {Warren~J.}\ \bibnamefont {Hehre}}, \bibinfo {author} {\bibfnamefont
  {Andreas}\ \bibnamefont {Klamt}}, \bibinfo {author} {\bibfnamefont {Henry
  F.~Schaefer}\ \bibnamefont {III}}, \bibinfo {author} {\bibfnamefont
  {Michael~W.}\ \bibnamefont {Schmidt}}, \bibinfo {author} {\bibfnamefont
  {C.~David}\ \bibnamefont {Sherrill}}, \bibinfo {author} {\bibfnamefont
  {Donald~G.}\ \bibnamefont {Truhlar}}, \bibinfo {author} {\bibfnamefont
  {Arieh}\ \bibnamefont {Warshel}}, \bibinfo {author} {\bibfnamefont {Xin}\
  \bibnamefont {Xu}}, \bibinfo {author} {\bibfnamefont {Al{\'a}n}\ \bibnamefont
  {Aspuru-Guzik}}, \bibinfo {author} {\bibfnamefont {Roi}\ \bibnamefont
  {Baer}}, \bibinfo {author} {\bibfnamefont {Alexis~T.}\ \bibnamefont {Bell}},
  \bibinfo {author} {\bibfnamefont {Nicholas~A.}\ \bibnamefont {Besley}},
  \bibinfo {author} {\bibfnamefont {Jeng-Da}\ \bibnamefont {Chai}}, \bibinfo
  {author} {\bibfnamefont {Andreas}\ \bibnamefont {Dreuw}}, \bibinfo {author}
  {\bibfnamefont {Barry~D.}\ \bibnamefont {Dunietz}}, \bibinfo {author}
  {\bibfnamefont {Thomas~R.}\ \bibnamefont {Furlani}}, \bibinfo {author}
  {\bibfnamefont {Steven~R.}\ \bibnamefont {Gwaltney}}, \bibinfo {author}
  {\bibfnamefont {Chao-Ping}\ \bibnamefont {Hsu}}, \bibinfo {author}
  {\bibfnamefont {Yousung}\ \bibnamefont {Jung}}, \bibinfo {author}
  {\bibfnamefont {Jing}\ \bibnamefont {Kong}}, \bibinfo {author} {\bibfnamefont
  {Daniel~S.}\ \bibnamefont {Lambrecht}}, \bibinfo {author} {\bibfnamefont
  {WanZhen}\ \bibnamefont {Liang}}, \bibinfo {author} {\bibfnamefont
  {Christian}\ \bibnamefont {Ochsenfeld}}, \bibinfo {author} {\bibfnamefont
  {Vitaly~A.}\ \bibnamefont {Rassolov}}, \bibinfo {author} {\bibfnamefont
  {Lyudmila~V.}\ \bibnamefont {Slipchenko}}, \bibinfo {author} {\bibfnamefont
  {Joseph~E.}\ \bibnamefont {Subotnik}}, \bibinfo {author} {\bibfnamefont
  {Troy~Van}\ \bibnamefont {Voorhis}}, \bibinfo {author} {\bibfnamefont
  {John~M.}\ \bibnamefont {Herbert}}, \bibinfo {author} {\bibfnamefont
  {Anna~I.}\ \bibnamefont {Krylov}}, \bibinfo {author} {\bibfnamefont
  {Peter~M.W.}\ \bibnamefont {Gill}}, \ and\ \bibinfo {author} {\bibfnamefont
  {Martin}\ \bibnamefont {Head-Gordon}},\ }\bibfield  {title} {\enquote
  {\bibinfo {title} {Advances in molecular quantum chemistry contained in the
  {Q-Chem} 4 program package},}\ }\href@noop {} {\bibfield  {journal} {\bibinfo
   {journal} {Mol. Phys.}\ }\textbf {\bibinfo {volume} {113}},\ \bibinfo
  {pages} {184--215} (\bibinfo {year} {2015})}\BibitemShut {NoStop}%
\bibitem [{\citenamefont {Li}\ \emph {et~al.}(2013)\citenamefont {Li},
  \citenamefont {Feng}, \citenamefont {Sun}, \citenamefont {Zhang},
  \citenamefont {Fan}, \citenamefont {Peterson}, \citenamefont {Xie},\ and\
  \citenamefont {III}}]{Li:2013}%
  \BibitemOpen
  \bibfield  {author} {\bibinfo {author} {\bibfnamefont {Huidong}\ \bibnamefont
  {Li}}, \bibinfo {author} {\bibfnamefont {Hao}\ \bibnamefont {Feng}}, \bibinfo
  {author} {\bibfnamefont {Weiguo}\ \bibnamefont {Sun}}, \bibinfo {author}
  {\bibfnamefont {Yi}~\bibnamefont {Zhang}}, \bibinfo {author} {\bibfnamefont
  {Qunchao}\ \bibnamefont {Fan}}, \bibinfo {author} {\bibfnamefont {Kirk~A.}\
  \bibnamefont {Peterson}}, \bibinfo {author} {\bibfnamefont {Yaoming}\
  \bibnamefont {Xie}}, \ and\ \bibinfo {author} {\bibfnamefont {Henry
  F.~Schaefer}\ \bibnamefont {III}},\ }\bibfield  {title} {\enquote {\bibinfo
  {title} {The alkaline earth dimer cations ({B}e$_2^+$, {M}g$_2^+$,
  {C}a$_2^+$, {S}r$_2^+$, and {B}a$_2^+$). {C}oupled cluster and full
  configuration interaction studies},}\ }\href {\doibase
  10.1080/00268976.2013.802818} {\bibfield  {journal} {\bibinfo  {journal}
  {Mol. Phys.}\ }\textbf {\bibinfo {volume} {111}},\ \bibinfo {pages}
  {2292--2298} (\bibinfo {year} {2013})}\BibitemShut {NoStop}%
\bibitem [{\citenamefont {Werner}\ \emph {et~al.}(2012)\citenamefont {Werner},
  \citenamefont {Knowles}, \citenamefont {Knizia}, \citenamefont {Manby},\ and\
  \citenamefont {Sch{\"u}tz}}]{molpro}%
  \BibitemOpen
  \bibfield  {author} {\bibinfo {author} {\bibfnamefont {H.-J.}\ \bibnamefont
  {Werner}}, \bibinfo {author} {\bibfnamefont {P.~J.}\ \bibnamefont {Knowles}},
  \bibinfo {author} {\bibfnamefont {G.}~\bibnamefont {Knizia}}, \bibinfo
  {author} {\bibfnamefont {F.~R.}\ \bibnamefont {Manby}}, \ and\ \bibinfo
  {author} {\bibfnamefont {M.}~\bibnamefont {Sch{\"u}tz}},\ }\bibfield  {title}
  {\enquote {\bibinfo {title} {Molpro: a general purpose quantum chemistry
  program package},}\ }\href@noop {} {\bibfield  {journal} {\bibinfo  {journal}
  {Comput. Mol. Sci.}\ }\textbf {\bibinfo {volume} {2}},\ \bibinfo {pages}
  {242--253} (\bibinfo {year} {2012})}\BibitemShut {NoStop}%
\bibitem [{\citenamefont {Ho}\ and\ \citenamefont {Rabitz}(1996)}]{ho:96}%
  \BibitemOpen
  \bibfield  {author} {\bibinfo {author} {\bibfnamefont {T.-S.}\ \bibnamefont
  {Ho}}\ and\ \bibinfo {author} {\bibfnamefont {H.}~\bibnamefont {Rabitz}},\
  }\bibfield  {title} {\enquote {\bibinfo {title} {A general method for
  constructing multidimensional molecular potential energy surfaces from {\em
  ab initio} calculations},}\ }\href@noop {} {\bibfield  {journal} {\bibinfo
  {journal} {J. Chem. Phys.}\ }\textbf {\bibinfo {volume} {104}},\ \bibinfo
  {pages} {2584} (\bibinfo {year} {1996})}\BibitemShut {NoStop}%
\bibitem [{\citenamefont {Presunka}\ and\ \citenamefont
  {Coxon}(1993)}]{Presunka:1993}%
  \BibitemOpen
  \bibfield  {author} {\bibinfo {author} {\bibfnamefont {P.I.}\ \bibnamefont
  {Presunka}}\ and\ \bibinfo {author} {\bibfnamefont {J.A.}\ \bibnamefont
  {Coxon}},\ }\bibfield  {title} {\enquote {\bibinfo {title} {High-resolution
  laser spectroscopy of excited bending vibrations (u2$\leq$2) of the
  \~{B}$^2{\Sigma}^+$ and \~{X}$^2{\Sigma}^+$ electronic states of {SrOH}:
  {A}nalysis of {\it l}-type doubling and {\it l}-type resonance},}\
  }\href@noop {} {\bibfield  {journal} {\bibinfo  {journal} {Can. J. Chem.}\
  }\textbf {\bibinfo {volume} {71}},\ \bibinfo {pages} {1689--1705} (\bibinfo
  {year} {1993})}\BibitemShut {NoStop}%
\bibitem [{\citenamefont {Oberlander}\ and\ \citenamefont
  {Parson}(1996)}]{Oberlander:1996}%
  \BibitemOpen
  \bibfield  {author} {\bibinfo {author} {\bibfnamefont {M.~D.}\ \bibnamefont
  {Oberlander}}\ and\ \bibinfo {author} {\bibfnamefont {J.~M.}\ \bibnamefont
  {Parson}},\ }\bibfield  {title} {\enquote {\bibinfo {title} {Laser excited
  fluorescence study of reactions of excited {Ca} and {Sr} with water and
  alcohols: {P}roduct selectivity and energy disposal},}\ }\href@noop {}
  {\bibfield  {journal} {\bibinfo  {journal} {J. Chem. Phys.}\ }\textbf
  {\bibinfo {volume} {105}},\ \bibinfo {pages} {5806--5816} (\bibinfo {year}
  {1996})}\BibitemShut {NoStop}%
\bibitem [{\citenamefont {van~der Hurk}\ \emph {et~al.}(1974)\citenamefont
  {van~der Hurk}, \citenamefont {Hollander},\ and\ \citenamefont
  {Alkemade}}]{SrOH_2}%
  \BibitemOpen
  \bibfield  {author} {\bibinfo {author} {\bibfnamefont {J.}~\bibnamefont
  {van~der Hurk}}, \bibinfo {author} {\bibfnamefont {Tj.}\ \bibnamefont
  {Hollander}}, \ and\ \bibinfo {author} {\bibfnamefont {C.Th.J.}\ \bibnamefont
  {Alkemade}},\ }\bibfield  {title} {\enquote {\bibinfo {title} {Excitation
  energies of strontium mono-hydroxide bands measured in flames},}\ }\href
  {\doibase https://doi.org/10.1016/0022-4073(74)90033-8} {\bibfield  {journal}
  {\bibinfo  {journal} {J. Quant. Spectrosc. and Radiat. Transf.}\ }\textbf
  {\bibinfo {volume} {14}},\ \bibinfo {pages} {1167 -- 1178} (\bibinfo {year}
  {1974})}\BibitemShut {NoStop}%
\bibitem [{\citenamefont {Cotton}\ and\ \citenamefont
  {Jenkins}(1968)}]{SrOH_3}%
  \BibitemOpen
  \bibfield  {author} {\bibinfo {author} {\bibfnamefont {D.~H.}\ \bibnamefont
  {Cotton}}\ and\ \bibinfo {author} {\bibfnamefont {D.~R.}\ \bibnamefont
  {Jenkins}},\ }\bibfield  {title} {\enquote {\bibinfo {title} {Dissociation
  energies of gaseous alkaline earth hydroxides},}\ }\href {\doibase
  10.1039/TF9686402988} {\bibfield  {journal} {\bibinfo  {journal} {Trans.
  Faraday Soc.}\ }\textbf {\bibinfo {volume} {64}},\ \bibinfo {pages}
  {2988--2997} (\bibinfo {year} {1968})}\BibitemShut {NoStop}%
\bibitem [{\citenamefont {Murad}(1981)}]{SrOH_5}%
  \BibitemOpen
  \bibfield  {author} {\bibinfo {author} {\bibfnamefont {Edmond}\ \bibnamefont
  {Murad}},\ }\bibfield  {title} {\enquote {\bibinfo {title} {Thermochemical
  properties of the gaseous alkaline earth monohydroxides},}\ }\href {\doibase
  10.1063/1.442567} {\bibfield  {journal} {\bibinfo  {journal} {J. Chem.
  Phys.}\ }\textbf {\bibinfo {volume} {75}},\ \bibinfo {pages} {4080--4085}
  (\bibinfo {year} {1981})}\BibitemShut {NoStop}%
\bibitem [{\citenamefont {Steimle}\ \emph {et~al.}(1992)\citenamefont
  {Steimle}, \citenamefont {Fletcher}, \citenamefont {Jung},\ and\
  \citenamefont {Scurlock}}]{Steimle1992}%
  \BibitemOpen
  \bibfield  {author} {\bibinfo {author} {\bibfnamefont {T.~C.}\ \bibnamefont
  {Steimle}}, \bibinfo {author} {\bibfnamefont {D.~A.}\ \bibnamefont
  {Fletcher}}, \bibinfo {author} {\bibfnamefont {K.~Y.}\ \bibnamefont {Jung}},
  \ and\ \bibinfo {author} {\bibfnamefont {C.~T.}\ \bibnamefont {Scurlock}},\
  }\bibfield  {title} {\enquote {\bibinfo {title} {A supersonic molecular beam
  optical stark study of caoh and sroh},}\ }\href {\doibase 10.1063/1.462007}
  {\bibfield  {journal} {\bibinfo  {journal} {J. Chem. Phys.}\ }\textbf
  {\bibinfo {volume} {96}},\ \bibinfo {pages} {2556--2564} (\bibinfo {year}
  {1992})}\BibitemShut {NoStop}%
\end{thebibliography}%

{\large \bf Supplementary Material}

\vspace*{3mm}

\begin{figure*}[h]
\includegraphics[width=0.7\textwidth]{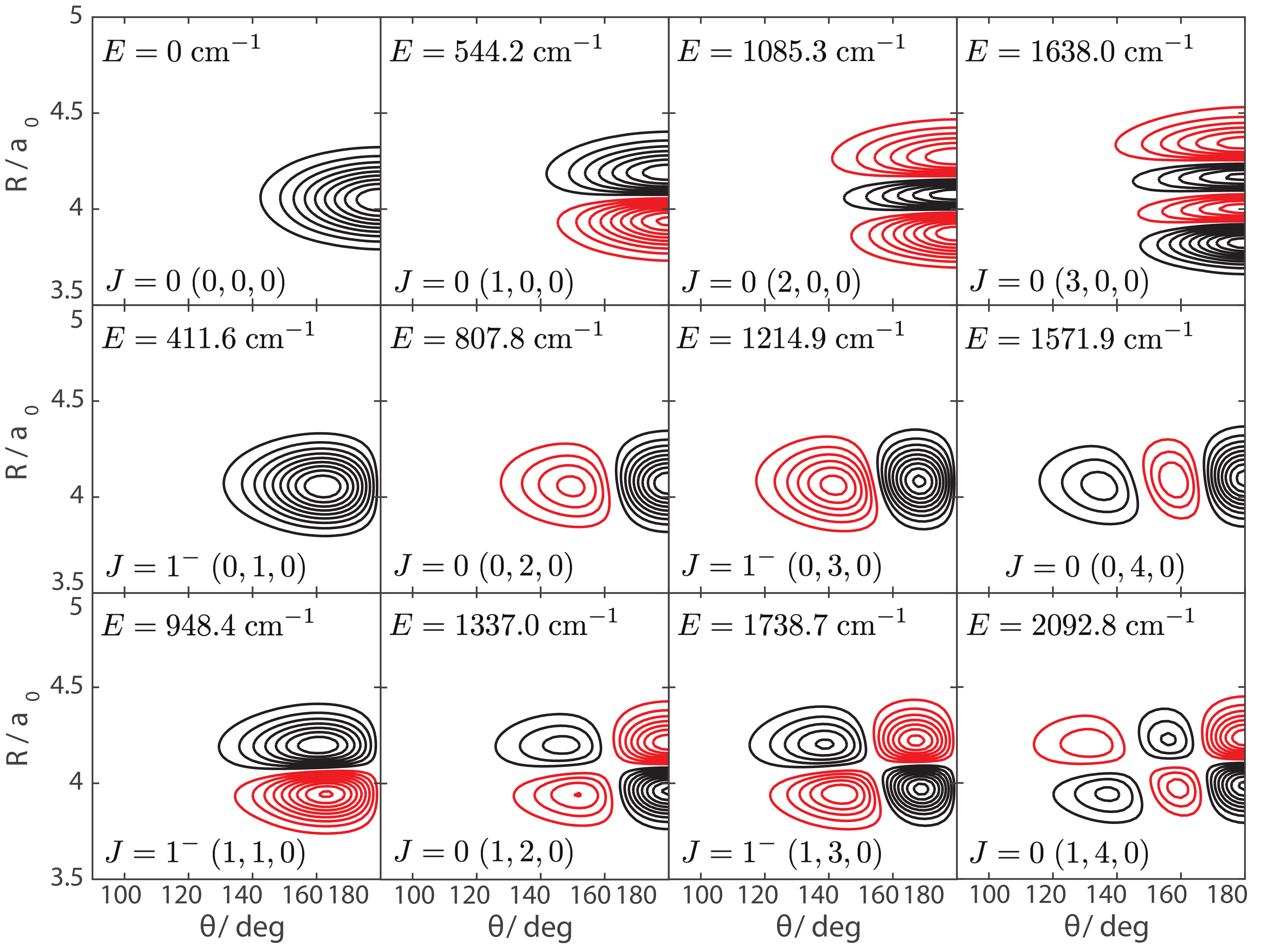}
\caption{
Contour plots of wavefunctions of $J^p$ ro-vibrational bound states 
as  functions of separation $R$ and angle $\theta$
for total angular momentum ${J=1}$ and  parity ${p=\pm1}$ of 
the $1\,^2\!A^{\prime}({\rm X}^2\Sigma^+)$ ground-state potential energy surface.  
Triplets $(v_{\rm s},v_{\rm b},v_{\rm OH}=0)$ indicate stretching 
and bending quantum numbers. Black and red lines denote wavefunction 
contours of opposite sign, respectively. Positive and negative 
parity states have an even and odd symmetry with respect to the 
reflection of angle $\theta\to 180^\circ-\theta$.
Energies in units of cm$^{-1}$ are given with respect to the 
$J^p=0^+$ $(0,0,0)$ level.
}
\label{fig:Ap_wfun}
\end{figure*}

\begin{figure*}
\includegraphics[width=0.7\textwidth]{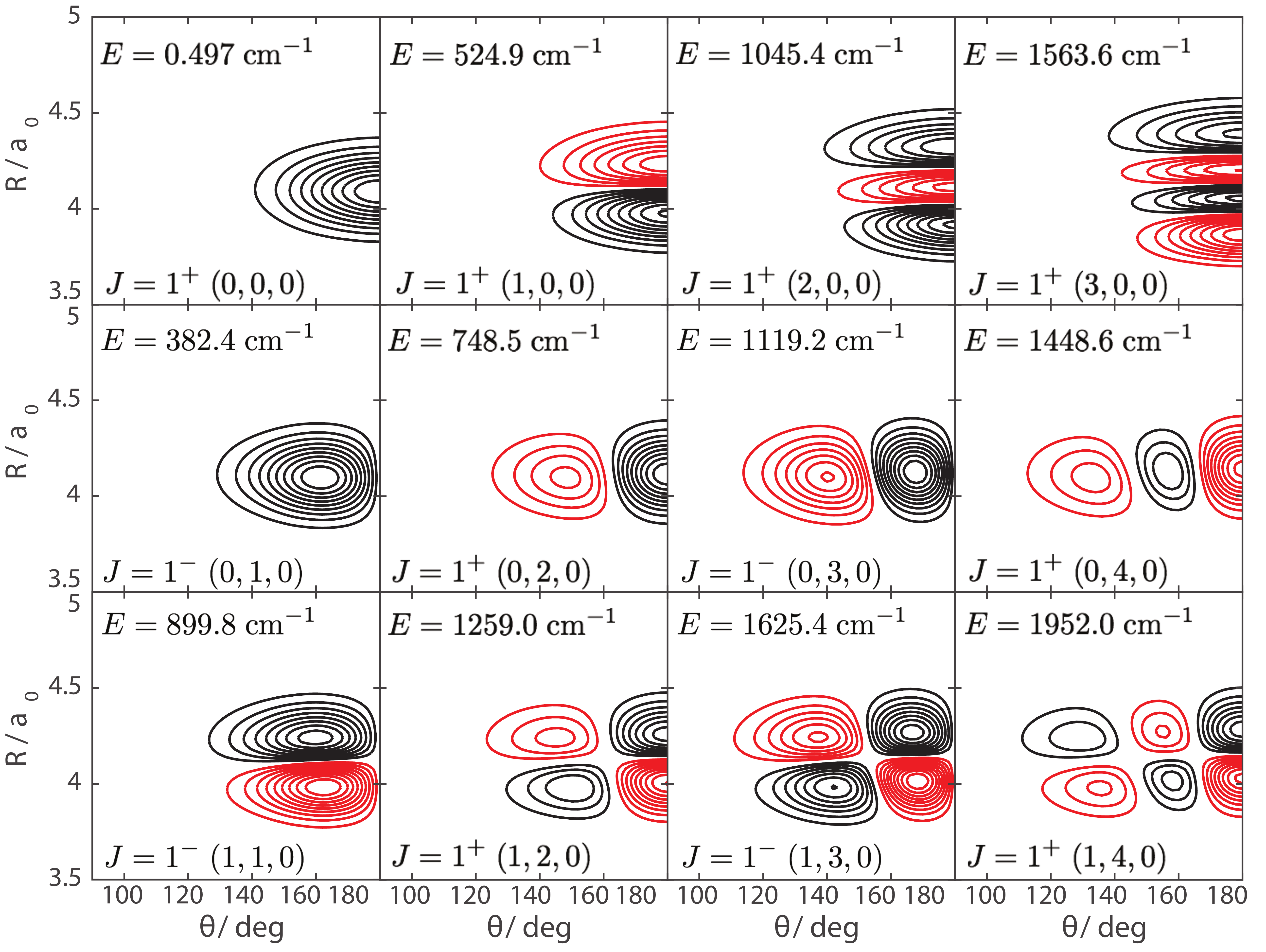}
\caption{
Contour plots of wavefunctions of rovibrational bound states as functions of separation $R$ and angle $\theta$ for 
$J^p=0^+$ and $1^\pm$ levels of the 
$1\,^2\!A^{\prime\prime}({\rm A}^2\Pi)$ potential energy surface. 
Energies in units of cm$^{-1}$ are given with respect to the 
$J^p=0^+$ $(0,0,0)$ level. 
}
\label{fig:App_wfun}
\end{figure*}

\begin{figure*} 
\includegraphics[width=0.7\textwidth]{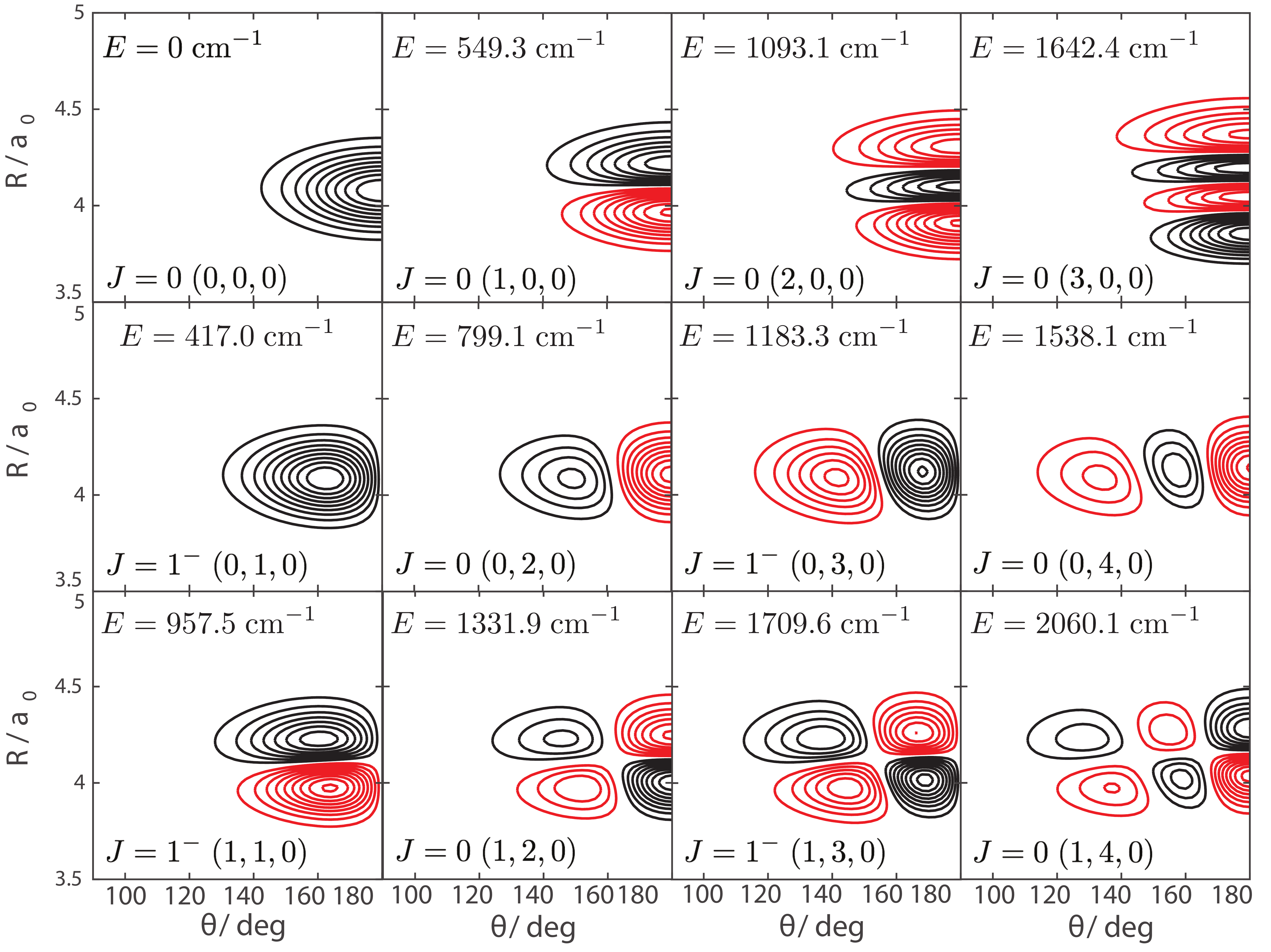}
\caption{
Contour plots of wavefunctions of rovibrational bound states as functions of separation $R$ and angle $\theta$ for 
$J^p=0^+$ and $1^{\pm}$ levels of  the 
$3\,^2\!A^{\prime}({\rm B}^2\Sigma^+)$ potential energy surface. 
Energies in units of cm$^{-1}$ are given with respect to the  
$J^p=0^+$ $(0,0,0)$ level.}
\label{fig:B_wfun}
\end{figure*}

\noindent
{\bf Supplementary Note 1: Stretching and bending modes in the ground and excited state potentials}. We have used a distributed Gaussian basis approach to obtain 
the energies and wavefunctions of the lowest few $J=0$ and 1 
stretching and bending modes in each of the 
$1\,^2\!A^{\prime}(X^2\Sigma^+)$, 
$1\,^2\!A^{\prime\prime}({\rm A}^2\Pi)$, and 
$3\,^2\!A^{\prime}({\rm B}^2\Sigma^+)$ PESs.
Supplementary Figures~\ref{fig:Ap_wfun}, \ref{fig:App_wfun}, and \ref{fig:B_wfun}
show contour plots of the real-valued vibrational wavefunctions 
for bending and stretching state progressions for these three PESs. 
One can identify bending and stretching quantum numbers, $v_{\rm b}$ 
and $v_{\rm s}$, by counting the lobes or studying  the nodal 
pattern of the wavefunctions in $\theta$ and $R$ directions, 
indicating that for the lowest energies the bending and stretching 
modes are not mixed. Positive and negative parity states have 
an even and odd symmetry with respect to the reflection 
$\theta\to 180^\circ-\theta$ and, thus, whether $v_{\rm b}$ is 
even or odd. Laser cooling is performed from the $J=1,p=+1$ $(0,0,0)$  
state of the  $1\,^2\!A^{\prime}(X^2\Sigma^+)$ potential.
We show this state in the left upper corner of Supplementary Figure~\ref{fig:Ap_wfun}. 
It lies approximately 0.5 cm$^{-1}$ above the absolute rovibrational 
ground state with $J=0$, corresponding to a $2B_0$ of 
excitation energy, where $B_0$ is a rotational constant of SrOH 
$1\,^2\!A^{\prime}(X^2\Sigma^+)$.

Franck-Condon factors of vibronic transitions, the square of
the overlap of vibrational wavefunctions, among the lowest bending 
and stretching modes of
$1\,^2\!A^{\prime}(X^2\Sigma^+)$, 
$1\,^2\!A^{\prime\prime}({\rm A}^2\Pi)$, and 
$3\,^2\!A^{\prime}({\rm B}^2\Sigma^+)$ are shown in 
Supplementary Tables~\ref{tab:FCFXA}, \ref{tab:FCFXB}, and \ref{tab:FCFAB}.



\begin{table*}
\caption{
Matrix of Franck-Condon factors (FCFs) between $(v_{\rm s},v_{\rm b},0)$  and 
$(v_{\rm s}',v_{\rm b}',0)$ SrOH rovibrational states of the 
excited $J=0$  $1\,^2\!A^{\prime\prime}({\rm A}^2\Pi)$ and 
ground $J=1$ $1\,^2\!A^{\prime}(X^2\Sigma^+)$ states, respectively. 
Only FCFs larger than $10^{-3}$ are shown.
}
\label{tab:FCFXA}
\begin{center}
\def\arraystretch{1.2}
\begin{tabular}{l|cccccccccccc}
\hline
A{\textbackslash}X  &   ~(000)~ & (100)~  &  (200)~ &   (300)~  &   (010)~  &  (020)~ &   (030) ~ &  (040)~  &  (110)~ &   (120)~  &   (130)~ &   (140)~
\Tstrut\Bstrut\\ 
\hline
(000)  &   {\bf 0.940} &  {\bf 0.058}    &    &     &      &      &      &      &      &      &  &  \Tstrut\\
(100)  &   {\bf 0.056} &     {\bf 0.829} &     {\bf 0.112}  &    {\bf 0.003} &       &      &      &      &     &       &      &     \\
(200)  &   {\bf 0.003}  &    {\bf 0.104} &     {\bf 0.736} &     {\bf 0.151} &       &      &      &      &     &       &      &     \\
(300)  &    &     {\bf 0.007} &    {\bf 0.136}  &   {\bf  0.640}  &       &      &      &      &     &       &      &    \\
(010)  &    &     &       &     &     {\bf 0.941}  &      &      &      &    {\bf 0.057} &       &      &     \\
(020)   &     &    &       &      &      &    {\bf 0.940}  &      &      &     &     {\bf 0.057}  &      &     \\
(030)   &    &      &      &      &      &      &    {\bf 0.942}  &      &      &      &    {\bf 0.054}  &     \\
(040)   &    &      &      &       &     &       &     &    {\bf 0.942}  &      &      &       &   {\bf 0.051} \\
(110)   &    &      &      &       &  {\bf 0.055}   &     &      &       &   {\bf 0.832}   &      &      &     \\
(120)   &     &      &      &       &     &   {\bf 0.055}  &      &       &     &    {\bf 0.832}   &     &     \\
(130)   &     &      &      &       &     &     &    {\bf 0.052}  &       &     &       &   {\bf 0.839}   &    \\
(140)   &     &      &      &       &     &     &      &    {\bf 0.049}   &    &       &       &  {\bf 0.844} \Bstrut\\ 
\hline
\end{tabular}
\end{center}
\end{table*}

\begin{table*}
\caption{
Matrix of Franck-Condon factors (FCFs)  between $(v_{\rm s},v_{\rm b},0)$  and 
$(v_{\rm s}',v_{\rm b}',0)$ SrOH rovibrational states of 
the excited $J=0$  $3\,^2\!A^{\prime}({\rm B}^2\Sigma^+)$ and 
ground $J=1$ $1\,^2\!A^{\prime}(X^2\Sigma^+)$ states, respectively. 
Only FCFs larger than $10^{-3}$ are shown.
}
\label{tab:FCFXB}
\begin{center}
\def\arraystretch{1.2}
\begin{tabular}{l|cccccccccccc}
\hline
B{\textbackslash}X    &   ~(000)~ &   (100)~ &  (200)~ &   (300)~ &   (010)~ &   (020)~ &   (030)~ &   (040)~ &   (110)~ &   (120)~ &   (130)~  &  (140)~
\Tstrut\Bstrut\\
\hline
(000) &   {\bf 0.993}  &   {\bf 0.006}   &    &     &     &     &     &     &     &     &      &   \Tstrut\\
(100) &   {\bf 0.006}  &   {\bf 0.981}   &  {\bf 0.011}  &     &     &     &     &     &     &     &      &   \\
(200) &     &   {\bf 0.011}   &  {\bf 0.969}  &   {\bf 0.017}  &     &     &     &     &     &     &      &   \\
(300) &     &      &  {\bf 0.017}  &   {\bf 0.962}  &     &     &     &     &     &     &      &   \\
(010) &     &      &    &     &   {\bf 0.994}  &     &    &     &   {\bf 0.004}  &     &      &   \\
(020) &     &      &    &     &     &   {\bf 0.994}  &     &    &     &   {\bf 0.003}  &      &   \\
(030) &     &      &    &     &     &     &   {\bf 0.992}  &     &     &     &   {\bf 0.003}   &   \\
(040) &     &      &    &     &     &     &     &   {\bf 0.996}  &     &     &      &    \\
(110) &     &      &    &     &   {\bf 0.004}  &     &     &     &   {\bf 0.984}  &     &      &   \\
(120) &     &      &    &     &     &   {\bf 0.003}  &     &     &     &   {\bf 0.987}  &      &   \\
(130) &     &      &    &     &     &     &   {\bf 0.002}  &     &     &     &   {\bf 0.979}   &  \\
(140) &     &      &    &     &     &     &     &      &     &     &      &  {\bf 0.982} \Bstrut\\ 
\hline
\end{tabular}
\end{center}
\end{table*}

\begin{table*}
\caption{
Matrix of Franck-Condon factors (FCFs)  between $(v_{\rm s},v_{\rm b},0)$  and 
$(v_{\rm s}',v_{\rm b}',0)$ SrOH rovibrational levels of the 
excited $J=0$  $1\,^2\!A^{\prime\prime}({\rm A}^2\Pi)$ and 
$J=0$  $ 3\,^2\!A^{\prime}({\rm B}^2\Sigma^+)$ states, respectively. Only FCFs larger than $10^{-3}$ are shown.
}
\label{tab:FCFAB}
\begin{center}
\def\arraystretch{1.2}
\begin{tabular}{l|cccccccccccc}
\hline
A{\textbackslash}B  &   ~(000)~ &   (100)~ &  (200)~ &   (300)~ &   (010)~ &   (020)~ &   (030)~ &   (040)~ &   (110)~ &   (120)~ &   (130)~  &  (140)~ 
\Tstrut\Bstrut\\
\hline
 (000) &    {\bf 0.969}  &   {\bf 0.030}  &         &          &          &          &          &          &          &          &           &      \Tstrut\\
 (100) &    {\bf 0.028}  &   {\bf 0.911}  &   {\bf 0.060} &          &          &          &          &          &          &          &           &         \\
 (200) &    {\bf 0.002}  &   {\bf 0.053}  &   {\bf 0.862} &    {\bf 0.081} &          &          &          &          &          &          &           &         \\
 (300) &           &   {\bf 0.004}  &   {\bf 0.067} &    {\bf 0.795} &          &          &          &          &          &          &           &         \\
 (010) &           &          &         &          &   {\bf 0.967} &          &          &          &   {\bf 0.033} &          &           &         \\
 (020) &           &          &         &          &          &   {\bf 0.963} &          &          &          &   {\bf 0.036} &           &         \\
 (030) &           &          &         &          &          &          &  {\bf  0.963} &          &          &          &   {\bf 0.035}  &         \\
 (040) &           &          &         &          &          &          &          &   {\bf 0.955} &          &   {\bf 0.002} &           &  {\bf 0.041} \\
 (110) &           &          &         &          &   {\bf 0.031} &          &  {\bf  0.002} &          &   {\bf 0.902} &          &           &         \\
 (120) &           &          &         &          &          &   {\bf 0.034} &          &   {\bf 0.002} &          &   {\bf 0.891} &           &         \\
 (130) &           &          &         &          &          &          &   {\bf 0.032} &          &          &          &    {\bf 0.887}  &         \\
 (140) &           &          &         &          &          &          &          &   {\bf 0.040} &          &          &           &   {\bf0.862} \Bstrut\\
\hline
\end{tabular}
\end{center}
\end{table*}


\end{document}